\documentclass[12pt]{article}
\usepackage{amssymb,amsmath,amsthm,mathtools}
\usepackage{dsfont}
\usepackage{graphicx}
\usepackage{natbib}
\usepackage[english]{babel}
\usepackage{booktabs}
\usepackage{enumitem}
\usepackage{xcolor}
\usepackage{tabularx}
\usepackage{multirow}
\usepackage{url}
\usepackage{caption,setspace}
\usepackage{varwidth}
\usepackage{newfloat}
\DeclareFloatingEnvironment[placement={hbt},name=List]{mylist}

\newcommand{\by}{\boldsymbol{y}}
\newcommand{\bY}{\boldsymbol{Y}}

\newcommand{\bU}{\boldsymbol{U}}
\newcommand{\bgamma}{\boldsymbol{\gamma}}

\newcommand{\bzeta}{\boldsymbol{\zeta}}
\newcommand{\bxi}{\boldsymbol{\xi}}
\newcommand{\btau}{\boldsymbol{\tau}}
\newcommand{\blambda}{\boldsymbol{\lambda}}

\newcommand{\balpha}{\boldsymbol{\alpha}}

\newcommand{\bA}{\boldsymbol{A}}
\newcommand{\bB}{\boldsymbol{B}}

\newcommand{\bD}{\boldsymbol{D}}
\newcommand{\bF}{\boldsymbol{F}}
\newcommand{\bG}{\boldsymbol{G}}
\newcommand{\bI}{\boldsymbol{I}}

\newcommand{\bs}{\boldsymbol{s}}
\newcommand{\bone}{\boldsymbol{1}}
\newcommand{\bzero}{\boldsymbol{0}}
\newcommand{\bP}{\boldsymbol{P}}
\newcommand{\bQ}{\boldsymbol{Q}}

\newcommand{\diff}{\mathop{}\!\mathrm{d}}
\newcommand{\indFun}[1]{\mathds{1}_{\{#1\}}}
\newcommand{\Pb}{\mathbb{P}}
\newcommand{\Ev}{\mathbb{E}}

\usepackage{scalerel,stackengine}

\title{\bf Stochastic Block Smooth Graphon Model}

\author{Benjamin Sischka and G\"{o}ran Kauermann}

\begin{document}

\def\spacingset#1{\renewcommand{\baselinestretch}%
{#1}\small\normalsize} \spacingset{1}

\maketitle
\let\thefootnote\relax\footnote{Benjamin Sischka is Research Assistant, Department of Statistics, Ludwig-Maximilians-Universit\"{a}t M\"{u}nchen, 80539 M\"{u}nchen, Germany (E-mail: \url{benjamin.sischka@stat.uni-muenchen.de}). G\"{o}ran Kauermann is Professor, Department of Statistics, Ludwig-Maximilians-Universit\"{a}t M\"{u}nchen, 80539 M\"{u}nchen, Germany (E-mail: \url{goeran.kauermann@stat.uni-muenchen.de}).
}

\bigskip
\begin{abstract}
	\noindent
	The paper proposes the combination of stochastic blockmodels with smooth graphon models. The first allow for partitioning the set of individuals in a network into blocks which represent groups of nodes that 
	presumably connect stochastically equivalently, 
	therefore often also called communities. Smooth graphon models instead assume that the network's nodes can be arranged on a one-dimensional scale such that closeness implies a similar connectivity behavior. 
	Both models belong to the model class of node-specific latent variables, entailing a natural relationship. While these model strands have developed more or less completely independently, this paper proposes their generalization towards \textit{stochastic block smooth graphon models}. This approach enables to exploit the advantages of both worlds. We pursue a general EM-type algorithm for estimation and demonstrate the usability 
	by applying the model to both simulated and real-world examples. 
\end{abstract}

\noindent%

\newpage
\spacingset{1.5}
\section{Introduction}
The statistical modeling of complex random networks has gained increasing interest over the last two decades and much development has taken place in this area. 
Data with 
network structure arise 
in many application fields and corresponding modeling frameworks are 
applied in sociology, biology, neuroscience, computer science, and others. To demonstrate the state of the art in statistical network data analysis, survey articles have been published by \cite{Goldenberg2009}, \cite{Snijders2011}, \cite{Hunter2012}, \cite{Fienberg2012}, and \cite{Salter-Townshend2012}. Moreover, monographs in this field are given by \cite{Kolaczyk2009}, \cite{Lusher2013}, \cite{Kolaczyk2014}, and \cite{Kolaczyk2017}. 

In order to capture the underlying structure within a given network, various modeling strategies based on different concepts have been developed. 
One very common model class in this context is given by the 
Node-Specific Latent Variable Models, 
see 
\cite{Matias2014} 
for an overview. The general concept in this broad model class is the assumption 
that, for a network of size $N$, the edges $Y_{ij}$, $i,j=1,\ldots,N$, between pairs of nodes 
can be modeled 
independently when conditioning on 
the node-specific latent quantities 
$\bxi_1, \ldots, \bxi_N$. 
To be precise, this 
generic 
model design 
can be formulated by independent Bernoulli random variables with appropriate success probabilities, i.e.
\begin{align}
Y_{ij} \mid \bxi_i,\bxi_j \stackrel{\text{ind.}}{\sim} \text{Bernoulli} (h(\bxi_i,\bxi_j)),
\label{eq:genMod}
\end{align}
where $0 \leq h(\cdot,\cdot) \leq 1$ refers to some overall connectivity pattern. 
This especially means that the connection probability for the node pair $(i,j)$ 
only depends on the corresponding quantities $\bxi_i$ and $\bxi_j$, which, depending on the model specification, are either considered as random variables themselves 
or simply as unknown but fixed parameters. In addition, note that $\bxi_i$ can generally be a multidimensional vector, although in many models it is simply used as a scalar. 
In case of undirected networks without self-loops, which is on what we are focusing in this work, the generating process (\ref{eq:genMod}) is only performed for $i<j$. 

This general framework includes several well-known models which are frequently used by practitioners 
in the field of statistical network analysis. The most poplar models of this class 
are the 
Stochastic Blockmodel 
(see \citealp{HollandLaskeyLein:83} or, for \textit{a posteriori} blockmodeling, \citealp{Snijders1997} and \citeyear{Nowicki2001}) as well as its variants (\citealp{Airoldi2008}, \citealp{Karrer2011}), the 
Latent Distance Model 
(Hoff and coauthors, 
\citeyear{Hoff2002}, \citeyear{Hoff2009}, \citeyear{Hoff2009a}, \citeyear{Hoff2020}, \citealp{Ma2020}), and the 
Graphon Model 
(Lov\'{a}sz, Borgs, and coauthors, \citeyear{Lovasz2006}, \citeyear{Borgs2007}, \citeyear{Borgs2010}, \citealp{diaconis2007graph}). Apparently, all these methods possess different abilities to cover diverse structural aspects in networks. 
In this line, for modeling a specific network at hand, it is often unknown what the requirements are in terms of structural expressiveness. 
In fact, it is commonly unclear which modeling strategy is best able to capture the present network structure. To detect the best method out of an ensemble of models and corresponding estimation algorithms, \cite{Li2020} recently developed a cross-validation procedure for model selection in the network context, see also \cite{Gao2020}. 
One step further, \cite{Ghasemian2020} and \cite{li2021} discuss the mixing of several model fits based on different weighting strategies. 

Although all node-specific latent variable models are more or less closely related, only little attention has been paid to 
a proper representation and integration of one model by another. As an advantage, such a merging potentially leads to a novel modeling approach, representing a combination of the respectively unified models. 
Steps into this direction have been taken by, for example, \citet[sec.\ 3]{Fosdick2019}, who developed a Latent Space Stochastic Blockmodel, where the within-community structure is modeled in the form of a latent distance model. 
In contrast, for constructing the Hierarchical Exponential Random Graph Model (HERGM), \cite{Schweinberger2015} combined the stochastic blockmodel with the Exponential Random Graph Model (ERGM), using it to uncover the within-community structure 
on the basis of subgraph frequencies. 
ERGMs themselves are generally beyond the formulations from (\ref{eq:genMod}). Instead, they focus on modeling the frequency of specific structural patterns. 

In this paper, 
we pick up the idea of model (\ref{eq:genMod}) but aim to estimate the connectivity pattern $h(\cdot,\cdot)$ 
in a more generalized way than previously developed concepts. 
To do so, we combine stochastic blockmodels with \textit{smooth} graphon models, 
leading to an extension that is able to capture the expressiveness of both models simultaneously. 
We will utilize previous results on smooth graphon estimation (\citealp{sischka2022based}) with EM-algorithm based stochastic blockmodel estimation (see e.g.\ \citealp{Daudin2008} or \citealp{DeNicola2021c}). The resulting model is flexible and the estimation routine is feasible for even large networks.

The rest of the paper is structured as follows. In Section~\ref{sec:hgm} we start with 
a discussion and literature review of both 
stochastic blockmodels and smooth graphon models, where we subsequently 
combine the two models, yielding a novel modeling approach. 
An EM-based estimation procedure for this new model is then developed in Section~\ref{sec:MCEM}, including the definition of a criterion for choosing the number of communities. 
In Section~\ref{sec:evalu}, its capability is demonstrated with reference to simulations, and we also show its applicability to real-world networks. 
The discussion and conclusion in Section~\ref{sec:discuss} completes the paper.

\section{Conceptualizing the Stochastic Block Smooth Gra\-phon Model}
\label{sec:hgm}

\subsection{The Stochastic Blockmodel}  %
In statistical network analysis, the stochastic blockmodel (SBM) is an extensively developed tool for modeling a clustering structure in networks, see \cite{Newman2006}, \cite{Choi2012b}, \cite{Peixoto2012}, \cite{Bickel2013}, and others. In its classical version, one assumes that each node $i=1,\ldots,N$ can be uniquely assigned to one of $K \in \mathbb{N}$ groups, where the probability of two nodes being connected then only depends on their group memberships. More precisely, the data-generating process can be formulated as drawing the node assignments $Z_i$ for all $i=1, \ldots, N$ independently from a categorical distribution given through
\begin{align}
\Pb(Z_i = k ; \balpha) = \alpha_k \geq 0 \quad \text{with } k = 1,\ldots,K, \text{ } 
\balpha = (\alpha_1,\ldots,\alpha_K) 
\text{ and } \sum_k \alpha_k = 1 ,
\label{eq:sbm1}
\end{align}
and, subsequently, simulating under conditional independence the edges through
\begin{align}
Y_{ij} \mid Z_i, Z_j \sim \text{Bernoulli}(p_{Z_i Z_j})
\label{eq:sbm2}
\end{align}
for $i < j$, where $Y_{ji} = Y_{ij}$ and $Y_{ii}=0$ by definition. 
In this formulation, $\balpha$ represents the vector of the (expected) group proportions and $p_{Z_i Z_j}$ is the corresponding entry of the edge probability matrix $\boldsymbol{P} = [p_{kl}]_{k,l = 1,\ldots,K} \in [0,1]^{K \times K}$. 
Referring to formulation (\ref{eq:genMod}), this construction is apparently equivalent to setting $\bxi_i = Z_i$ and $h(Z_i,Z_j) = p_{Z_i Z_j}$. 
Moreover, this modeling approach can also be viewed as a mixture of Erd\H{o}s-R\'{e}nyi-Gilbert models 
(\citealp{Daudin2008}) since the connections between all pairs of nodes from two particular communities (or also within one community) are described as stochastically independent and having 
the same probability. 

Although the model formulation is straightforward, the estimation is rather complex because both the latent community memberships and the model parameters need to be estimated. 
The literature of this \textit{a posterior} blockmodeling starts with the works of Snijders and Nowicki (\citeyear{Snijders1997}, \citeyear{Nowicki2001}) and since then has been elaborated extensively 
(\citealp{Handcock2007}, \citealp{Decelle2011}, \citealp{Rohe2011}, \citealp{Choi2012b}, \citealp{Peixoto2017}, 
and others). 
As an additional complication, usually also the number of communities has to be inferred from the data. 
Works focusing on that issue are, among others, \cite{Wang2017}, \cite{Chen2018}, \cite{Newman2016}, and \cite{Riolo2017}. 

\subsection{The Smooth Graphon Model}
Another modeling approach which makes use of latent quantities to capture complex network structures is the 
graphon model. 
In contrast to the SBM, the latent variables in the graphon model are scaled continuously on $[0,1]$, but again the connectivity is assumed to depend only on those latent quantities. The data-generating process induced 
by the graphon model can more precisely be formulated as follows. First, the latent quantities are independently drawn from a uniform distribution, i.e.
\begin{align}
U_i \stackrel{\text{i.i.d.}}{\sim} \text{Uniform} (0,1).
\label{eq:uniform}
\end{align}
Secondly, the network entries are sampled conditionally independently in the form of
\begin{align}
Y_{ij} \mid U_i, U_j \sim \text{Bernoulli} (w(U_i, U_j)),
\label{eq:graphon}
\end{align}
for $i < j$, 
where again $Y_{ji} = Y_{ij}$ and $Y_{ii}=0$. The bivariate function $w: [0,1]^2 \rightarrow [0,1]$ is the so-called graphon. Choosing $\bxi_i = U_i$ and $h(U_i, U_j) = w(U_i, U_j)$ yields again the 
representation 
in the form of (\ref{eq:genMod}). 
In comparison with the SBM, the graphon model does usually \textit{not} decompose a network into groups of equally behaving actors. Instead, it allows for a more flexible structure. 
Comparing the respective connectivity objects, the flexibility and thus the complexity of the graphon, $w(\cdot,\cdot)$, is far higher than it is for the edge probability matrix, $\boldsymbol{P}$. 
In fact, 
to get this complexity under control with regard to estimation, 
some additional constraints are required. Commonly one assumes that $w(\cdot,\cdot)$ is smooth, meaning that it fulfills some H\"older or Lipschitz condition (\citealp{WolfeOl:14}, \citealp{Gao2015}, and \citealp{Klopp2017}). We call such a model a \textit{smooth} graphon model (SGM). 
With respect to this smoothness assumption, many works apply a histogram estimator, 
see e.g.\ \cite{wolfe2013nonparametric}, \cite{Airoldi2013}, \cite{Chan2014}, or \cite{Yang2014}. 
\cite{sischka2022based} make use of (linear) B-spline regression to guarantee a smooth and stable estimation of $w(\cdot, \cdot)$. 
In contrast, some other works make less restrictive assumptions but solely aim for estimating the edge probabilities $\Pb (Y_{ij} = 1 \mid U_i, U_j)$ rather than $w(\cdot,\cdot)$ itself, see \cite{Chatterjee2015} and \cite{Zhang2017}.

\subsection{The Stochastic Block Smooth Graphon Model}
\label{sec:sbgm}
\begin{figure}%
	\centering
	\begin{minipage}[t]{0.49\textwidth}
		\centering
		\includegraphics[width=1\textwidth]{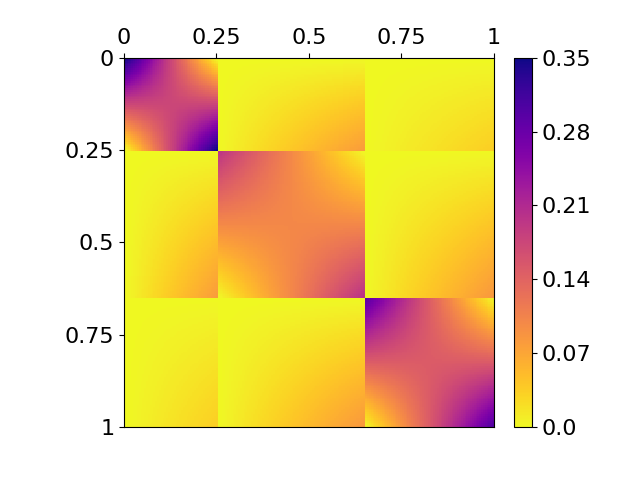}
	\end{minipage}
	\hfill
	\begin{minipage}[t]{0.49\textwidth}
		\centering
		\includegraphics[width=1\textwidth]{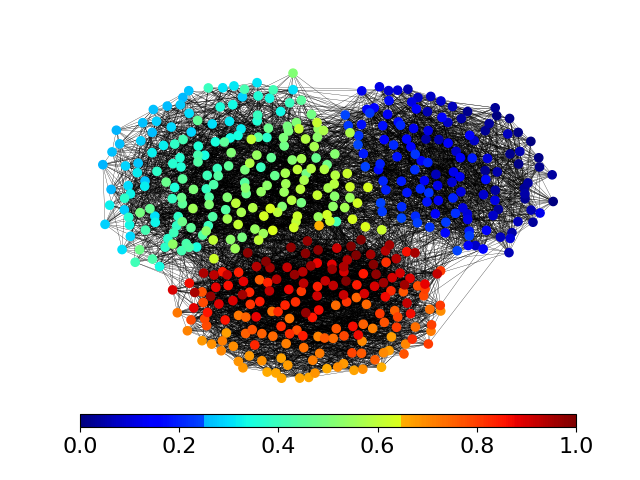}  %
	\end{minipage}
	\caption{Exemplary stochastic block smooth graphon model $w(\cdot,\cdot)$ with 3 communities, represented as heat map (left). A simulated network of size $500$ which is based on this model is given on the right, with node coloring referring to the sampled $U_i$. This network exhibits a clear community structure (global division) but also smooth transitions within the communities (local structure).}
	\label{fig:hier-graphon}
\end{figure}

Both the SBM and the SGM 
are build on 
underlying assumptions which appear to be restrictive conditions---namely strict homogeneity within the communities and 
overall smoothness, 
respectively. 
Thus, we purse 
to create a new model class which does not suffer from such limitations. 
To do so, we combine the two modeling approaches towards what we call a Stochastic Block Smooth Graphon Model (\mbox{SBSGM}). To be specific, we assume the node assignments $Z_i$ to be drawn from (\ref{eq:sbm1}) and draw independently $U_i$ from (\ref{eq:uniform}) for $i=1,\ldots,N$. Then (\ref{eq:sbm2}) and (\ref{eq:graphon}) are replaced by 
\begin{align}
Y_{ij} \mid Z_i, Z_j, U_i , U_j \stackrel{\text{i.i.d.}}{\sim} \text{Bernoulli}(\tilde{w}_{Z_i Z_j}(U_i,U_j)),
\label{eq:sbgm1}
\end{align}
where, for each pair of blocks and also within blocks, connectivity is now formulated by an individual smooth graphon $\tilde{w}_{kl}(\cdot,\cdot)$, $k,l=1,\ldots,K$. 
Apparently, if $K=1$ we obtain an SGM, 
while constant $\tilde{w}_{kl}(\cdot,\cdot)$ with values $p_{kl}$ leads to an SBM. 
This model can be reformulated in a simplified form by combining the node assignments (\ref{eq:sbm1}) and the latent quantities (\ref{eq:uniform}) in the following way. We draw $U_i$ from (\ref{eq:uniform}) and, given $U_i$ for $i=1,\ldots,N$, we formulate for $i < j$
\begin{align}
Y_{ij} \mid U_i , U_j \stackrel{\text{i.i.d.}}{\sim} \text{Bernoulli}(w_{\bzeta} (U_i,U_j)) ,
\label{eq:sbgm2}
\end{align}
where $w_{\bzeta} (\cdot,\cdot)$ is a partitioned graphon which is smooth within the blocks spanned by 
$\bzeta= (\zeta_0=0,\zeta_1, \ldots,\zeta_K=1)$, meaning within 
$(\zeta_{k - 1}, \zeta_{k}) \times (\zeta_{l -1}, \zeta_{l})$ for $k,l = 1,\ldots,K$. To be precise, to transform the formulation from (\ref{eq:sbgm1}) to (\ref{eq:sbgm2}), we set $\zeta_k = \sum_{l=1}^k \alpha_l$ and
\[
	w_{\bzeta} (u,v) = \tilde{w}_{k_u k_v} \left( \frac{u - \zeta_{k_u -1}}{\zeta_{k_u} - \zeta_{k_u -1}} , \frac{v - \zeta_{k_v -1}}{\zeta_{k_v} - \zeta_{k_v -1}} \right),
\]
where $k_u \in \{1,\ldots,K\}$ is given such that $\zeta_{k_u -1} \leq u < \zeta_{k_u}$, i.e.\ $k_u = \sum_k \indFun{u \geq \zeta_k}$. 
We also here remain with the common convention of symmetry ($Y_{ji}=Y_{ij}$) and the absence of self-loops ($Y_{ii}=0$). An exemplary \mbox{SBSGM} together with a corresponding simulated network is illustrated in Figure~\ref{fig:hier-graphon}. 
As a special property in terms of expressiveness, this model allows for smooth local structures under a global division into groups. 

Note that the assumption of such a piecewise smooth structure in the context of graphon models has also been proposed before, see e.g.\ \cite{Airoldi2013} or \cite{Zhang2017}. Nonetheless, there is a major conceptional distinction in the modeling perspective pursued here. 
While in previous works, lines of discontinuity were \textit{merely allowed}, we now \textit{explicitly incorporate} them as structural breaks. We stress that this novel modeling approach---which also rules the estimation---has a strong impact on 
uncovering the network's underlying structure. 
This is demonstrated in the simulation studies in Section~\ref{sec:simStud}, where the true structure including structural breaks can be fully recovered. 
In this regard, we also refer to \cite{li2021}, who showed that mixing the estimation results of graphon models with those of SBMs yields an improvement in the goodness of fit.

\subsection{Piecewise Smoothness and Semiparametric Model Formulation}  %
\begin{figure}%
	\centering
	\includegraphics[width=1\textwidth]{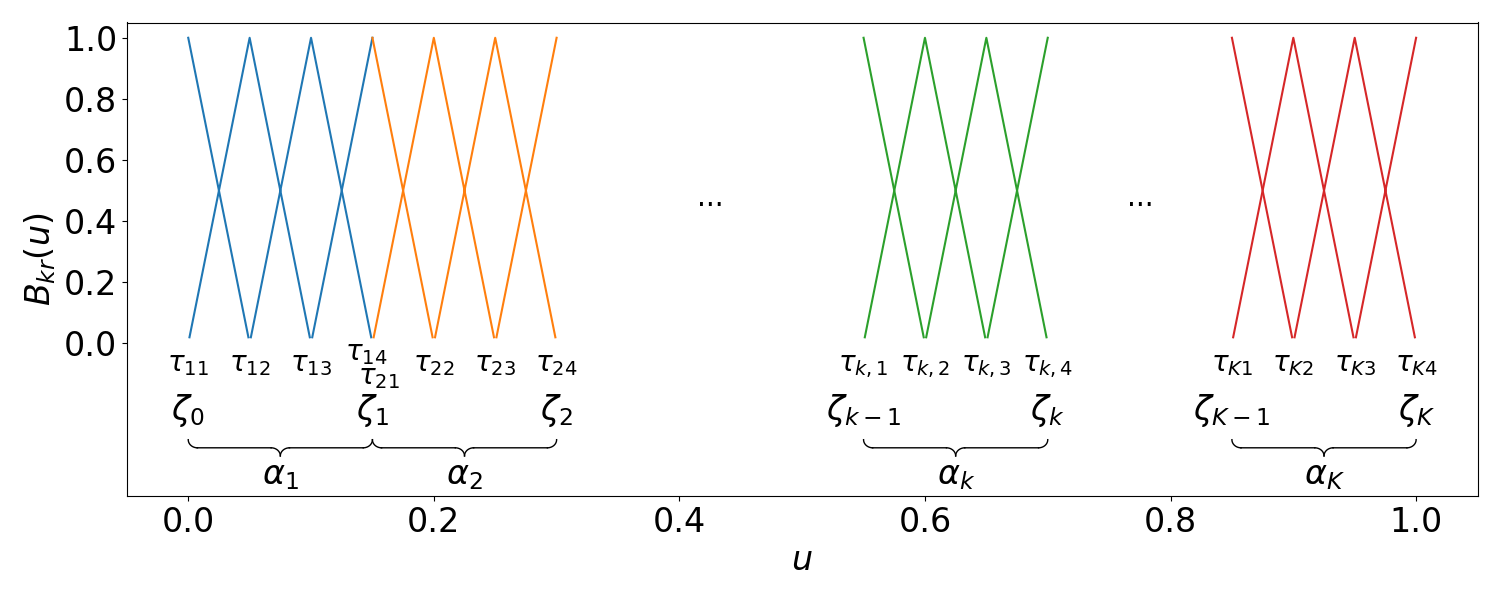}
	\caption{Disjoint univariate linear B-spline bases, colored in blue, orange, green, and red, respectively. Applying the tensor product yields the basis to construct blockwise independent B-spline functions for approaching a stochastic block smooth graphon model. Note that this illustration shows the special case of equal community proportions.}
	\label{fig:splineBase}
\end{figure}

\label{sec:piecesmooth}
In general, we define the \mbox{SBSGM} to be specified by a piecewise Lipschitz graphon with lines of discontinuity. 
In this context, a graphon $w(\cdot ,\cdot )$ satisfies piecewise the Lipschitz condition if there exist boundaries $0= \zeta_0 < \zeta_1 < \ldots < \zeta_{K} = 1$ and a constant $M \geq 0$ such that for all $u,u^\prime \in (\zeta_{k-1}, \zeta_k)$, $v, v^\prime \in (\zeta_{l-1}, \zeta_l)$
\begin{align}
	\vert w(u,v) - w(u^\prime,v^\prime) \vert \leq M \Vert (u,v)^\top - (u^\prime,v^\prime)^\top \Vert 
	\label{eq:lip}
\end{align}
for any $k,l = 1, \ldots, K$, where $\Vert \cdot \Vert$ is the Euclidean norm. We indicate this in the notation 
by making use of the subscript $\bzeta$, meaning that $w_{\bzeta} (\cdot,\cdot)$ is piecewise Lipschitz continuous with corresponding boundaries $\bzeta$. 
In the case of $M=0$, this implies the representation of an SBM. 
To achieve a semiparametric structure from this theoretical model formulation, we follow the approach of \cite{sischka2022based} and make use of linear B-splines to approximate and estimate the (local) smooth structures. Here, we extend the overall smooth representation to the 
piecewise smooth format. 
To do so, we construct a mixture of 
B-splines, 
i.e.\ we formulate blockwise B-spline functions on disjoint bases in the form of
\begin{align}
	w_{\bzeta,\bgamma}^{\text{\textit{spline}}}(u,v) = \sum_k \sum_l \indFun{\zeta_{k-1} \leq u < \zeta_k} \indFun{\zeta_{l-1} \leq v < \zeta_l} [\bB_k (u) \otimes \bB_l (v)] \bgamma_{kl},
	\label{eq:hierBspline}
\end{align}
where $\otimes$ is the Kronecker product and $\bB_k(\cdot) = (B_{k1}(\cdot),\ldots,B_{k L_k}(\cdot)) \in \mathbb{R}^{1 \times L_k}$ is a linear B-spline basis on $[\zeta_{k-1}, \zeta_k]$, normalized to have maximum value 1. We refer to Figure~\ref{fig:splineBase} for a color-coded exemplification. 
In the above formulations, the component boundaries 
$\bzeta = (\zeta_0,\ldots,\zeta_K)$ 
are specified through (\ref{eq:lip}) 
and the inner knots of the $k$-th one-dimensional B-spline component with length $L_k$ are denoted by 
$\btau_k=(\tau_{k1}, \ldots, \tau_{kL_k})$, where $\tau_{k1} = \zeta_{k-1}$ and $\tau_{kL_k} = \zeta_{k}$. 
Moreover, 
$\btau= (\btau_1, \ldots, \btau_K)$ denotes the overall vector of B-spline knots 
and, 
for given separated bases $\bB_k(\cdot)$, $k=1,\ldots,K$, the parameter vector is given in the form of 
\begin{align*}
\bgamma =(\bgamma_{11}^\top, \ldots, \bgamma_{1K}^\top, \bgamma_{21}^\top, \ldots, \bgamma_{KK}^\top )^\top
\end{align*}
with $\bgamma_{kl} = (\gamma_{kl,11}, \ldots, \gamma_{kl,1L_l}, \gamma_{kl,21}, \ldots, \gamma_{kl,L_k L_l})^\top$. 
This piecewise spline representation then serves as suitable approximation of the \mbox{SBSGM} and, apparently, increasing $L_k$ reduces the 
approximation error
\[
	\sqrt{\iint \left\vert w_{\bzeta}(u,v) - w_{\bzeta,\bgamma}^{\text{\textit{spline}}}(u,v) \right\vert^2 \diff u \diff v}.
\]
For an adequate representation, we further choose the inner knots $\btau = (\tau_{11},\ldots,\tau_{1 L_1}, \allowbreak \tau_{21},\ldots,\tau_{K L_K})$ to be distributed among the community segments in proportion to the group sizes. To be precise, let $L = \vert \btau \vert$ be the overall amount of inner knots. We then locate the inner knots so that they are as equidistant as possible. To achieve this, we define the number of knots within the single communities, $L_k \in \mathbb{N}$, $k=1,\ldots,K$, 
through minimizing 
the sum over the relative downward deviation, i.e.\ through
\[
	\min_{L_1, \ldots, L_K} \sum_k \left[ \max \left\{\frac{\left(\zeta_k - \zeta_{k-1}\right) \cdot L}{L_k}, 1 \right\} -1 \right]
\]
with respect to $\sum_k L_k = L$ and $L_k \geq 2$ for $k=1,\ldots,K$. Subsequently, the knots for community $k$, $\btau_k=(\tau_{k1}, \ldots, \tau_{kL_k})$, are placed equidistantly within the community segment $[\zeta_{k-1}, \zeta_k]$, 
which is finally used for the B-spline formulation (\ref{eq:hierBspline}). 

The above formulations allow to apply penalized B-spline regression readily. 
The capability 
of such an approach as well as the general role of penalized semiparametric modeling concepts is discussed, for example, by \cite{Eilers1996}, \cite{Wood2017}, \cite{Ruppert2003}, and \cite{Kauermann2011}. 
For the sake of simplicity, we subsequently drop the superscript \textit{spline} in the notation whenever it is clear from the context that the formulation refers to a spline representation.

\subsection{The Identifiability Issue}
\label{sec:identIs}
As discussed above, the \mbox{SBSGM} describes an explicit specification of a graphon model. 
As such, it 
also suffers from non-identifiability. To be precise, \cite{diaconis2007graph} showed that two graphons 
$w(\cdot, \cdot)$ and $w^\prime (\cdot, \cdot)$ 
describe the same network generating process if and only if there exist two measure-preserving functions $\varphi, \, \varphi^\prime: [0,1] \rightarrow [0,1]$ such that 
\begin{align}
w(\varphi(u),\varphi(v))= w^\prime (\varphi^\prime(u),\varphi^\prime(v))
\label{eq:identGraphon}
\end{align} 
for almost all $(u,v)^\top \in [0,1]^2$. 
To circumvent this identifiability issue and to guarantee uniqueness, some papers have postulated that 
\begin{align}
g(u) = \int w(u,v) \diff v
\label{eq:gMono}
\end{align}
is strictly increasing, see e.g.\ \cite{Bickel2009} or \cite{Chan2014}. This, however, is a strong restriction on the generality of the 
graphon model. To give an example, it excludes the model with $w(u, v)=(u v)^{2}+((1-u)(1-v))^{2}$ 
since there exists no measurable-preserving function $\varphi: [0,1] \rightarrow [0,1]$ such that $w(\varphi(\cdot), \varphi(\cdot))$ is well-defined and fulfills condition (\ref{eq:gMono}). 
We therefore avoid to employ such a restrictive uniqueness assumption. 
Instead, we emphasize that identifiability issues such as label switching are an inherent problem in all mixture models (see e.g.\ \citealp{Stephens2000}), which can often be handled through appropriate estimation routines. 
A further discussion on this issue, including conditions that allow us to derive a proper estimate, is given in the Supplementary Material.

\section{EM-type Algorithm}  %
\label{sec:MCEM}

For fitting the \mbox{SBSGM} to a given network, 
the latent positions $U_1,\ldots,U_N$ and the parameters $(\bzeta, \bgamma)$ need to be estimated simultaneously. This is a typical task for an EM-type algorithm, which 
aims at deriving information about 
the unknown quantities in an iterative way. 
Regarding the inherent community structure, we assume the number of groups, $K \in \mathbb{N}$, as given for now. A discussion on that issue is provided in Section~\ref{sec:number}.

\subsection{The MCMC-E-Step}
\label{subsec:estep}
The conditional distribution of $\bU$ given $\by$ is rather complex 
and hence calculating the expectation cannot be solved analytically. Therefore, we apply MCMC techniques for carrying out the E-step. In that regard, the full-conditional distribution of $U_i$ can be formulated as
\begin{align}
f(u_i \mid u_1, \ldots, u_{i-1}, u_{i+1}, \ldots, u_N, \by) \propto \prod_{j \neq i} w_{\bzeta}(u_i, u_j)^{y_{ij}} (1-w_{\bzeta}(u_i, u_j))^{1-y_{ij}}. 
\label{eq:Gibb1}
\end{align}
Based on that, we can construct a Gibbs sampler, which allows consecutive drawings for $U_1,\ldots,U_N$. 
For its concrete implementation, we replace $w_{\bzeta}(\cdot,\cdot)$ by its current estimate. 
Finally, we derive reliable means for the node positions by appropriately summarizing the MCMC sequence. 
Technical details are provided in Section~\ref{AppSec:gibbs} of the Appendix. 

\begin{figure}%
	\centering
	\includegraphics[width=1\textwidth,trim={0 1cm 0 0},clip]{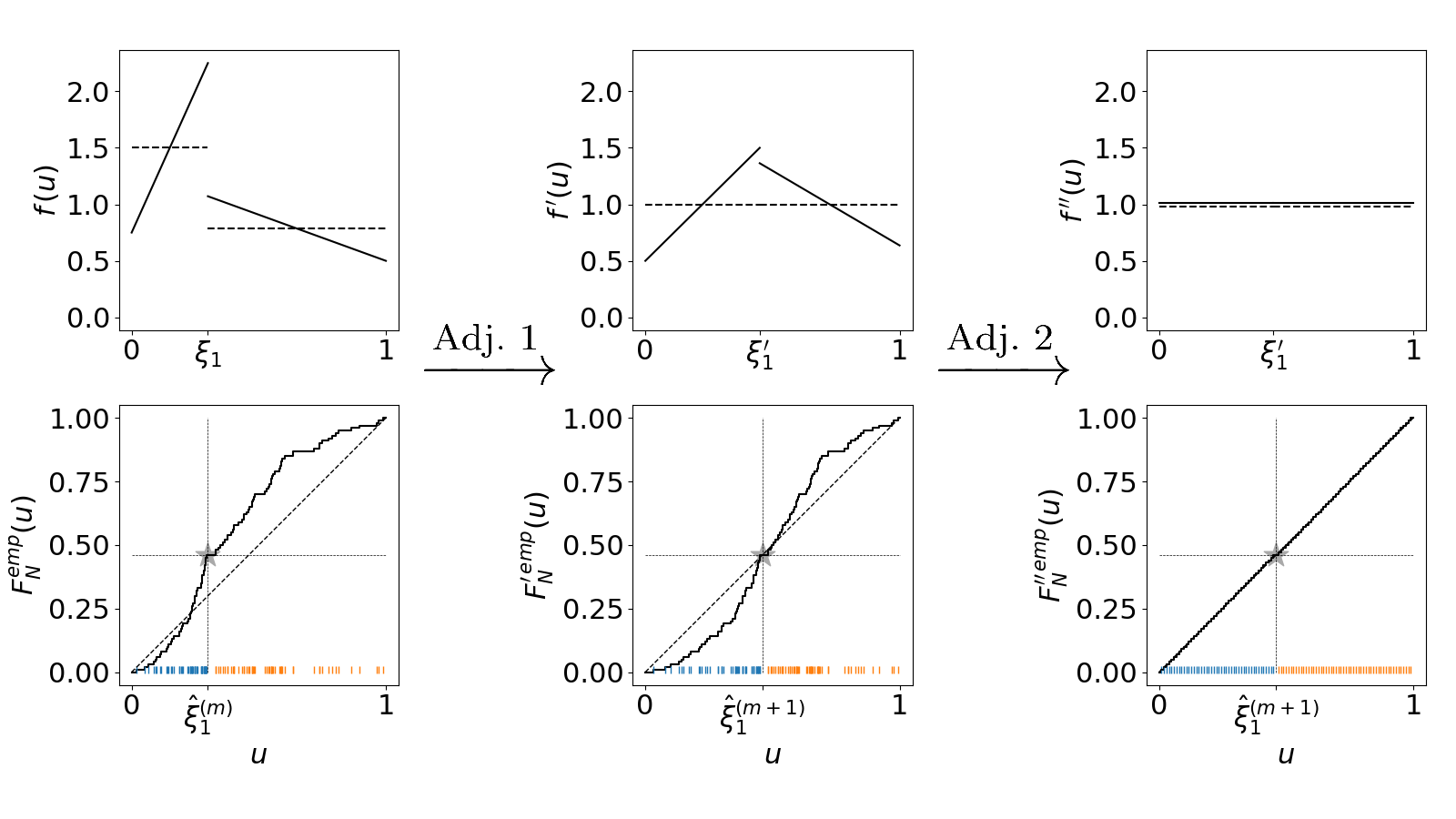}  %
	\caption{
		Adjustment of the latent quantities' distribution and the community boundaries. 
		Top: Three distributions for the latent quantities $U_i$ which are equivalent in terms of representing the same data-generating process (under applying transformation (\ref{eq:equiv}) to 
		$w_{\bzeta}(\cdot,\cdot)$ 
		accordingly). The solid line represents the density $f(u)$, while the dashed line illustrates the 
		frequency density over the communities, 
		i.e.\ $\Pb(U \in [\zeta_{k-1}, \zeta_k)) / (\zeta_k - \zeta_{k-1})$.
		Bottom: Implementation of the adjustment in the algorithm with regard to the empirical cumulative distribution function (including the realizations of $U_1, \ldots, U_N$ as vertical bars at the bottom). 
		The gray star illustrates the community boundary in comparison with the proportion of the two communities.
	}
	\label{fig:Corr12}
\end{figure}

We are however faced with an additional identifiability issue, 
which we want to motivate as follows. 
Assume first an \mbox{SBSGM} as in 
(\ref{eq:sbgm2}), but 
allow the distribution of the latent quantities $U_i$, subsequently denoted by $F(\cdot)$, to be not necessarily uniform but arbitrarily continuous instead. 
In this context, note that for any  
strictly increasing 
continuous 
transformation $\varphi^\prime: [0,1] \rightarrow [0,1]$, we have that with $U_i^\prime = \varphi^\prime(U_i)$, 
meaning $F^\prime(\cdot) \equiv F({\varphi^\prime}^{-1}(\cdot))$, 
we obtain an equivalent \mbox{SBSGM} through
\begin{align}
	w^\prime_{\bzeta^\prime} (u^\prime, v^\prime) = w_{\bzeta}({\varphi^\prime}^{-1} (u^\prime), {\varphi^\prime}^{-1} (v^\prime))
	\label{eq:equiv}
\end{align}
for $(u^\prime, v^\prime)^\top \in [0,1]^2$. 
In comparison with formulation (\ref{eq:identGraphon}), here $\varphi^\prime(\cdot)$ 
implies a modification of the probability measure on the domain $[0,1]$ and therefore it is \textit{no} measure-preserving transformation (except for the identity map $\varphi^\prime(u) = u$). 
In this regard, 
we can transform any 
``unregularized'' 
\mbox{SBSGM} 
with continuous distribution $F(\cdot)$ 
into a 
``regularized'' 
\mbox{SBSGM} 
with uniform distribution 
by applying $\varphi^\prime(U_i) := F(U_i)$. 
Nonetheless, 
given that the two models 
($F(\cdot)$, $w_{\bzeta}(\cdot,\cdot)$) and ($F^\prime(\cdot)$, $w^\prime_{\bzeta^\prime}(\cdot,\cdot)$) 
are not distinguishable in terms of the probability mass function induced on a network, 
this involves an additional identifiability issue which needs to be 
handled 
post hoc 
in the estimation routine. 
We tackle this issue 
by making use of two separate transformations $\varphi_1^\prime, \varphi_2^\prime: [0,1] \rightarrow [0,1]$, 
adjusting between and within groups, respectively. 
This is sketched in Figure~\ref{fig:Corr12}, where we demonstrate both the theoretical realization and the concrete  implementation in the algorithm. 
Considering the three theoretical distributions in the upper row, when combined with (\ref{eq:equiv}) above they can be thought of as equivalent. 
At that, the transformation from the left to the middle distribution refers to adjusting the size of a community 
according to its probability mass. 
More precisely, by relocating the boundaries $\zeta_k$ 
in the form of $\zeta_k^\prime = \varphi_1^\prime (\zeta_k)$ 
we can achieve a size-proportional distribution between the groups, meaning that $\Pb(U_i^\prime \in [\zeta_{k-1}^\prime, \zeta_k^\prime)) = \zeta_k^\prime - \zeta_{k-1}^\prime$ with $U_i^\prime = \varphi_1^\prime(U_i)$. We define this as Adjustment~1 and, in fact, Adjustment~1 
results through 
the M-step by estimating the block sizes. 
The middle distribution, however, still exhibits a second problem, 
namely 
non-uniformity within the groups. 
To solve this, we make use of a second transformation $\varphi_2^\prime (\cdot)$ 
to achieve an overall uniform distribution. 
We label this as Adjustment~2. 

The concrete implementation in the algorithm of both adjustments is sketched in the lower row of Figure~\ref{fig:Corr12}. Moreover, this is described in detail in Section~\ref{AppSec:gibbs} of the Appendix. 
We denote the final result of the E-step in the $m$-th iteration, i.e.\ the outcome achieved through Gibbs sampling and applying Adjustment~1 and Adjustment~2, by $\hat{\bU}^{\prime\prime (m)} = (\hat{U}_1^{\prime\prime (m)}, \ldots, \hat{U}_N^{\prime\prime (m)})$.

\subsection{The M-Step}
\label{subsec:splines}

\subsubsection{Linear B-Spline Regression}
In consequence of representing the \mbox{SBSGM} as a mixture of (linear) B-splines as in (\ref{eq:hierBspline}), we are now able to view the estimation as semiparametric regression problem, which can be solved by a regular maximum likelihood approach. 
Given the spline formulation, 
the full log-likelihood 
results in
\begin{multline*}
	\ell(\bgamma) %
	= \sum\limits_{\substack{i,j \\ j \neq i}} \sum_{k,l} \indFun{\hat{\zeta}_{k-1}^{(m+1)} \leq \hat{U}^{\prime\prime (m)}_i < \hat{\zeta}^{(m+1)}_k} \indFun{\hat{\zeta}^{(m+1)}_{l-1} \leq \hat{U}^{\prime\prime (m)}_j < \hat{\zeta}^{(m+1)}_l} \\
	\cdot \left[ y_{ij} \, \log \left( \bB^{(m+1)}_{kl,ij} \bgamma_{kl} \right) + \left( 1- y_{ij} \right) \, \log \left( 1 - \bB^{(m+1)}_{kl,ij} \bgamma_{kl} \right) \right], 
\end{multline*}
where $\bB^{(m+1)}_{kl,ij} = \bB^{(m+1)}_{k}(\hat{U}_i^{\prime\prime (m)}) \otimes \bB^{(m+1)}_{l}(\hat{U}_j^{\prime\prime (m)})$ and $\bB^{(m+1)}_{k}(\cdot)$ is the B-spline basis on $[\hat{\zeta}_{k-1}^{(m+1)}, \hat{\zeta}^{(m+1)}_k]$. Taking the derivative leads to the score function
\begin{multline*}
	\bs (\bgamma) = \sum\limits_{\substack{i,j \\ j \neq i}} \sum_{k,l} \indFun{\hat{\zeta}^{(m+1)}_{k-1} \leq \hat{U}_i^{\prime\prime (m)} < \hat{\zeta}^{(m+1)}_k} \indFun{\hat{\zeta}^{(m+1)}_{l-1} \leq \hat{U}_j^{\prime\prime (m)} < \hat{\zeta}^{(m+1)}_l} \\
	\cdot {\bB^{(m+1)}_{kl,ij}}^\top \left( \frac{y_{ij}}{w_{\hat{\bzeta}^{(m+1)}, \bgamma}(\hat{U}_i^{\prime \prime (m)},\hat{U}_j^{\prime\prime (m)})} - \frac{1-y_{ij}}{1-w_{\hat{\bzeta}^{(m+1)}, \bgamma}(\hat{U}_i^{\prime\prime (m)},\hat{U}_j^{\prime\prime (m)})} \right).
\end{multline*}
Moreover, taking the expected second order derivative 
gives us the Fisher matrix
\begin{multline*}
	\bF(\bgamma) = \sum\limits_{\substack{i,j \\ j \neq i}} \sum_{k,l} \indFun{\hat{\zeta}^{(m+1)}_{k-1} \leq \hat{U}_i^{\prime\prime (m)} < \hat{\zeta}^{(m+1)}_k} \indFun{\hat{\zeta}^{(m+1)}_{l-1} \leq \hat{U}_j^{\prime\prime (m)} < \hat{\zeta}^{(m+1)}_l} \\
	\cdot {\bB^{(m+1)}_{kl,ij}}^\top \bB_{kl,ij}^{(m+1)} \left[ w_{\hat{\bzeta}^{(m+1)}, \bgamma} \left( \hat{U}_i^{\prime\prime (m)},\hat{U}_j^{\prime\prime (m)} \right) \cdot \left( 1 - w_{\hat{\bzeta}^{(m+1)}, \bgamma} \left( \hat{U}_i^{\prime\prime (m)},\hat{U}_j^{\prime\prime (m)} \right) \right) \right]^{-1}.
\end{multline*}
We now aim to maximize $\ell(\bgamma)$, 
which usually could be done by Fisher scoring. However, we additionally need to ensure that the resulting estimate $\hat{w}^{(m+1)} (\cdot,\cdot) := w_{\hat{\bzeta}^{(m+1)}, \hat{\bgamma}^{(m+1)}}(\cdot,\cdot)$ in the 
\mbox{$(m+1)$-th} EM iteration fulfills symmetry and boundedness, which is why we impose additional (linear) side constraints on $\bgamma$. 
To guarantee symmetry, we accommodate $\gamma_{kl,pq} = \gamma_{lk,qp}$ for all $k,l \in \{1,\ldots,K\}$ and $p \neq q$. Moreover, the condition of $\hat{w}^{(m+1)}(\cdot,\cdot)$ being bounded to $[0,1]$ can be formulated as $0 \leq \gamma_{kl,pq} \leq 1$. Therefore, both side constraints can be incorporated in the linear forms of 
$\bG \bgamma \geq (\bzero^\top, -\bone^\top)^\top$ and $\bA \bgamma = \bzero$ for matrices $\bG$ and $\bA$ chosen accordingly, where $\bzero = (0,\ldots,0)^\top$ and $\bone = (1,\ldots,1)^\top$ are of corresponding sizes. Hence, maximizing $\ell(\bgamma)$ with respect to the postulated side constraints can be considered as an (iterated) quadratic programming problem, which can be solved using standard software (see e.g.\ \citealp{cvxopt} or \citealp{quadprog}).

\subsubsection{Penalized Estimation}
\label{sec:pen}
Following the motivation and idea underlying the penalized spline estimation (see \citeauthor{Eilers1996}, \citeyear{Eilers1996} or \citeauthor{Ruppert2009}, \citeyear{Ruppert2009}), we additionally impose a penalty on the coefficients to achieve smoothness. This is necessary since we intend to choose the 
overall dimension $L$ of the mixture of B-splines 
to be large and unpenalized estimation will lead to wiggled estimates. 
Apparently, the relation between components of the mixture will be left unpenalized and we only want to induce smoothness within the components. 
To do so, we penalize the difference between ``neighboring'' elements of $\bgamma_{kl}$. Let therefore
\[
\bD_{k} = \begin{pmatrix}
1 & -1 & \phantom{-}0 & \phantom{-}\ldots & \phantom{-}0 \\
0 & \phantom{-}1 & -1 & \phantom{-}\ldots & \phantom{-}0 \\
\vdots & \multicolumn{3}{c}{\ddots} & \phantom{-}\vdots \\
0 & \phantom{-}\ldots & \phantom{-}0 & \phantom{-}1 & -1 \\
\end{pmatrix} \in \mathbb{R}^{(L_k-1) \times L_k}
\]
be the first order difference matrix. We then penalize $\left[ \bD_{k} \otimes \bI_l \right] \bgamma_{kl}$ and $\left[ \bI_k \otimes \bD_{l} \right] \bgamma_{kl}$, where $\bI_k$ is the identity matrix of size $L_k$. This leads to the penalized log-likelihood
\[
	\ell^p (\bgamma, \blambda) = \ell (\bgamma) - \frac{1}{2} \bgamma^\top \bQ_{\blambda} \bgamma ,
\]
where $\bQ_{\blambda}$ is the diagonal matrix $\text{diag}\{ \lambda_{11} \bQ_{11}, \ldots, \lambda_{1K} \bQ_{1K}, \lambda_{21} \bQ_{21}, \allowbreak \ldots, \lambda_{KK} \bQ_{KK} \}$ with $\quad \bQ_{kl} = \left( \bD_{k} \otimes \bI_l \right)^{\top} \left( \bD_{k} \otimes \bI_l \right) +  \left( \bI_k \otimes \bD_{l} \right)^{\top} \left( \bI_k \otimes \bD_{l} \right)$ 
and $\blambda = (\lambda_{11}, \ldots, \lambda_{1K}, \lambda_{21}, \allowbreak \ldots, \lambda_{KK})$ 
serving as vector of smoothing parameters for the respective blocks. 
In this configuration, the resulting estimate apparently depends on the penalty parameter vector $\blambda$. 
Setting $\lambda_{kl} \rightarrow 0$ for $k,l=1,\ldots,K$ yields an unpenalized fit, while setting $\lambda_{kl} \rightarrow \infty$ leads to a piecewise constant \mbox{SBSGM}, i.e.\ an 
SBM. 
Therefore, the smoothing parameter vector $\blambda$ needs to be chosen 
in a data-driven way. 
For example, this can be realized by relying on the Akaike Information Criterion (AIC) (see \citealp{Hurvich1989} or \citealp{Burnham2002}). In the present context, this can be formulated as
\begin{align}
	\text{{AIC}} (\blambda) &= - 2 \, \ell (\hat{\bgamma}^p) + 2 \, \operatorname{df}(\blambda), %
	\label{eq:AIC}
\end{align}
where $\hat{\bgamma}^p$ is the penalized parameter estimate and $\operatorname{df}(\blambda)$ represents the cumulated degrees of freedom within the blocks. 
We define the latter in the common way as the trace of the product of the inverse penalized Fisher matrix ${[\bF^{p}]}^{-1} ( \hat{\bgamma}^p, \blambda )$ and the unpenalized Fisher matrix, see 
\citet[page 211 and the following pages]{Wood2017}. 
To be precise, we define
\begin{align*}
	\operatorname{df}(\blambda) = \operatorname{tr} \left\{ {[\bF^{p}]}^{-1} ( \hat{\bgamma}^p, \blambda ) \bF ( \hat{\bgamma}^p ) \right\} 
\end{align*}
with $\operatorname{tr}\{\cdot\}$ as the trace of a matrix. 
Making use of $\operatorname{df}_{kl}(\lambda_{kl}) = {[\bF_{kl}^{p}]}^{-1} ( \hat{\bgamma}_{kl}^p, \lambda_{kl} ) \bF_{kl} ( \hat{\bgamma}_{kl}^p )$ with $\bF_{kl} ( \hat{\bgamma}_{kl}^p)$ being the submatrix of $\bF ( \hat{\bgamma}^p )$ which refers to the subvector $\hat{\bgamma}_{kl}^{p}$ and for the penalized fisher matrix equivalently, this calculation can be reduced to 
$
	\operatorname{df}(\blambda) = \sum_{k,l} \operatorname{df}_{kl}(\lambda_{kl})
$ 
since ${[\bF^{p}]}^{-1} ( \hat{\bgamma}^p, \blambda )$ and $\bF ( \hat{\bgamma}^p)$ are both block diagonal matrices. 
Applying this simplification, 
we can rephrase (\ref{eq:AIC}) to
\begin{align}
	\text{{AIC}} (\blambda) &= \sum_{k,l} \left\{ - 2 \, \ell_{kl} (\hat{\bgamma}_{kl}^p)+ 2 \, \operatorname{df}_{kl}(\lambda_{kl}) \right\},
	\label{eq:AIC2}
\end{align}
where $\ell_{kl} (\cdot)$ is the partial likelihood of all potential connections falling into the \mbox{$(k,l)$-th} component. 
This representation allows us to optimize for $\lambda_{kl}$ separately. Following this procedure finally leads us to parameter estimate $\hat{\bgamma}^{(m+1)}$ in the $(m+1)$-th iteration of the EM algorithm.

\subsection{Choice of the Number of Communities}  %
\label{sec:number}
In real-world networks, the number of communities, $K$, is usually unknown. 
Preferably, this should also be inferred from the data. 
We pursue this by following two different intuitions, which we subsequently combine to an appropriate model selection criterion. 

On the one hand, 
it seems plausible to adopt 
methods for determining the number of communities in the SBM context. 
A common approach 
to do so 
is given by the Integrated Classification Likelihood ($\operatorname{ICL}$) criterion (\citealp{Daudin2008}, \citealp{Come2015}, \citealp{Mariadassou2010}). 
However, the more complex structure in the \mbox{SBSGM} needs to be observed since a higher flexibility within communities can to some extent compensate for too few groups and vice versa. 

As an alternative approach, we here exploit the already formulated AIC from (\ref{eq:AIC}), extending it towards a model selection strategy with respect to $K$. 
In fact, we propose to combine the AIC with a Bayesian Information Criterion (BIC). That is, we select the smoothing parameters using the AIC as described above, but for the number of blocks we impose a stronger penalty by replacing the factor $2$ in an extended AIC with the logarithmized sample size. We consider this to be in line with \cite{Burnham2004}, who conclude that the AIC is more reliable when the ground truth can be described through many tapering effects (smooth within-community differences), whereas the BIC should be preferred under
the presence of a few big effects only (number of groups). 

In order to formulate the BIC part, 
we first need to think carefully how the model complexity grows with increasing number of groups and what the corresponding sample size is.
The degrees of freedom originating from the number of groups 
comprises two aspects, 
the 
$K-1$ boundary parameters $\zeta_1, \ldots, \zeta_{K-1}$ and the 
$K^2$ 
basis connectivity parameters between and within communities (comparable to $\bP$ in the SBM context). 
As number of observations we propose to take $N$ (number of nodes) for the boundary parameters and 
$N(N-1)$ 
(number of edges) for the connectivity parameters. 
Moreover, we have to take into account that the second component in 
(\ref{eq:AIC}) 
already contains the degrees of freedom 
that are induced by 
the basis connectivity parameters. 
This can be easily seen by setting $\lambda_{kl} \rightarrow \infty$, leading to $\operatorname{df}(\blambda) = K^2$. 
Thus, this quantity needs to be subtracted from $\operatorname{df}(\blambda)$, 
what, however, has no effect on the optimization with respect to $\blambda$. 
Putting all together, we propose to extend (\ref{eq:AIC}) towards the complete 
model selection criterion
\begin{align}
	- 2 \, \ell (\hat{\bzeta}_K, \hat{\bgamma}_K)                                       + 2 \, \{ \operatorname{df}(\hat{\blambda}_K) - K^2 \} + \log\{ N(N-1) \} \, K^2 + \log \{ N\} \, (K-1),
	\label{eq:modelSelect}
\end{align}
where $\hat{\bzeta}_K$, $\hat{\bgamma}_K$, and $\hat{\blambda}_K$ are the final estimates according to the above EM procedure for given $K$, which is also indicated by the subscript. 
We emphasize that criterion (\ref{eq:modelSelect}) is equivalent to the ICL up to the different parameterization of the log-likelihood and the term 
$2 \, \{ \operatorname{df}(\hat{\blambda}_K) - K^2\}$ 
for penalizing the additional smooth differences within communities. 
Moreover, in case that the smoothing parameters $\lambda_{kl}$ are set to infinity, the criterion reduces exactly to the ICL for SBMs.

\section{Application}  %
\label{sec:evalu}
We examine the performance of our approach for 
both simulated and real-world networks. 
For an ``uninformative'' implementation, 
we initialize the algorithm by using a random permutation of $(i/(N+1) : \, i = 1,\ldots, N)$ as a starting estimate for the latent quantities. 
At the same time, we place the community boundaries equidistantly within $[0,1]$, i.e.\ we set $\hat{\zeta}_{k}^{(0)} = k/K$ for $k=0,\ldots,K$. Since different initializations might lead to different final results, we repeat the estimation procedure with different random permutations for $\hat{\bU}^{(0)}$ and choose the best outcome. 

If one aims at cutting computational costs, also ``informative'' initializations are conceivable. 
Reasonable starting values for $\bU$ could exemplary be derived through applying multidimensional scaling to the nodes' connectivity, i.e.\ $\by_{i \bullet} = (y_{i1}, \ldots, y_{iN})$, employing the reduction to one dimension. This follows the intuition of the \mbox{SBSGM}, according to which (per block) nearby nodes behave similarly. In this framework, $\bzeta$ can be initialized by determining the largest gaps within $\hat{\bU}^{(0)}$ or the highest differences between connectivity after ordering the nodes accordingly. 
However, if the focus is on finding the best result, as in our case, we recommend repeating the algorithm with different random initializations.

\subsection{Synthetic Networks}
\label{sec:simStud}
In the scenario of simulations, we order the final estimate according to the ground-truth model with respect to both, the arrangement of the groups and the within-group orientation. 
That is, applying $\varphi: [0,1] \rightarrow [0,1]$ from (\ref{eq:identGraphon}) %
to either swap communities or to reverse the arrangement within a group from back to front. 
Note that both however does not affect the actual estimation result and only helps to make illustrations more comparable. 

\subsubsection{Assortative Structures with Smooth Within-Group Differences}  %
\begin{figure}%
	\centering	
	\begin{minipage}[t]{0.49\textwidth}
		\centering
		\includegraphics[width=1\textwidth]{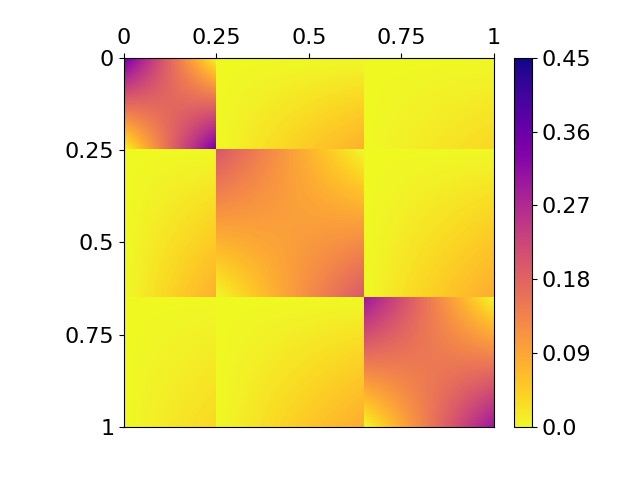}
	\end{minipage}
	\hfill
	\begin{minipage}[t]{0.49\textwidth}
		\centering
		\includegraphics[width=1\textwidth]{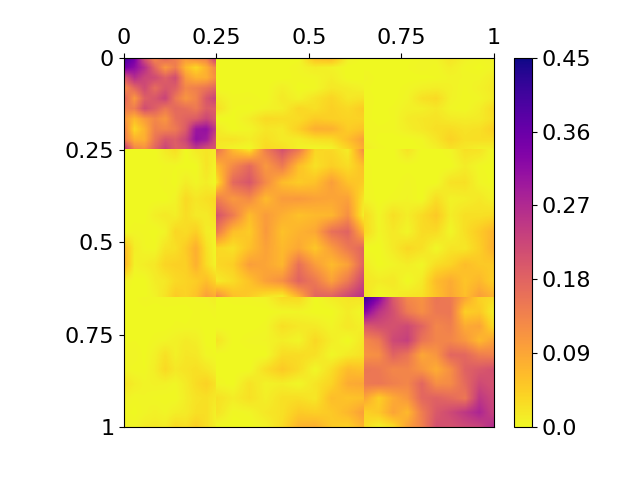}
	\end{minipage}
	\vfill
	
	\begin{minipage}[t]{0.49\textwidth}
		\centering
		\includegraphics[width=1\textwidth,trim={0 0 0 1.1cm},clip]{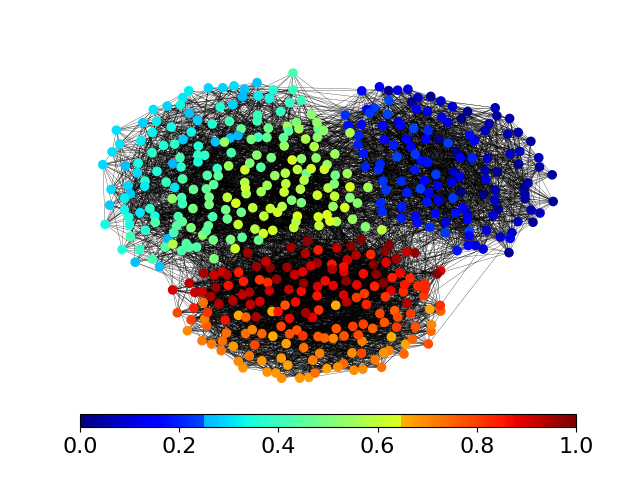}
	\end{minipage}
	\hfill
	\begin{minipage}[t]{0.49\textwidth}
		\centering
		\vspace*{-5.5cm}
		\hspace*{-.67cm}
		\includegraphics[width=1\textwidth]{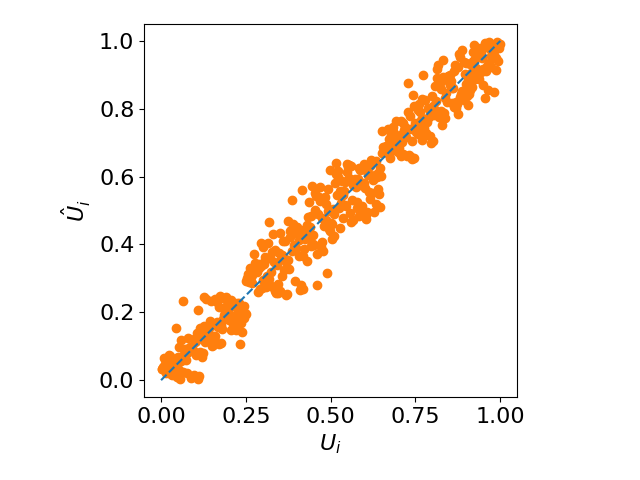}
	\end{minipage}
	\caption{Estimation results for the synthetic \mbox{SBSGM} from Figure~\ref{fig:hier-graphon} (shown again at top left, rescaled according to the estimate's range from $0$ to $0.45$). The top right plot shows the final \mbox{SBSGM} estimate, i.e.\ after convergence of the algorithm. The estimation is based on the simulated network of size $N = 500$ at the bottom left, where nodes are colored according to $\hat{U}_i \in [0,1]$. A comparison between the true simulated 
	$U_i$ 
	and the estimates 
	$\hat{U}_i$ 
	is illustrated at the bottom right.}
	\label{fig:synGraphon}
\end{figure}

\label{sec:assortNet}
To showcase the general applicability of our method, we at first consider again the \mbox{SBSGM} from Figure~\ref{fig:hier-graphon}. 
Starting with determining the number of groups, the first row of Table~\ref{tab:crit} 
shows the corresponding values for criterion (\ref{eq:modelSelect}). 
\begin{table}%
	\begin{center}
		\begin{tabular*}{\textwidth}{l @{\extracolsep{\fill}} cccccc}
			\toprule
			$K$                                        & 1		& 2		& 3		& 4		& 5		& 6 		\\
			\midrule \midrule
			Assortative network	                     & \multirow{2}{*}{8.960} 	& \multirow{2}{*}{8.978} & \multirow{2}{*}{\textbf{8.953}} & \multirow{2}{*}{8.956}		& \multirow{2}{*}{8.963}		& \multirow{2}{*}{8.981} 		\\
			(see Section~\ref{sec:assortNet})	     &  					& 					&  & 		& 		&  		 \\[0.2cm]
			Core-periphery network 	                     & \multirow{2}{*}{\textbf{9.799}} 					& \multirow{2}{*}{9.804} 					& \multirow{2}{*}{9.825} & \multirow{2}{*}{9.833}		& \multirow{2}{*}{9.843}		& \multirow{2}{*}{9.846} 		 \\
			(Section~\ref{sec:hubNet})        &  					& 				&  & 		& 		&  		 \\[0.2cm]
			Network with differing 	     & \multirow{3}{*}{\textbf{10.287}} 					& \multirow{3}{*}{10.330} 					& \multirow{3}{*}{10.394} & \multirow{3}{*}{10.373}		& \multirow{3}{*}{10.402}		& \multirow{3}{*}{10.422} 		 \\
			preferences 	     &  					&  					&  & 		& 		&  		 \\
			(Section~\ref{sec:diffPrefNet})        &  					&  					&  & 		& 		&  		 \\[0.2cm]
			Political blogs 		                 & \multirow{2}{*}{18.937} 					    & \multirow{2}{*}{\textbf{18.356}}				& \multirow{2}{*}{18.464} & \multirow{2}{*}{18.533}		& \multirow{2}{*}{19.091}		& \multirow{2}{*}{18.966} 		 \\ 
			(Section~\ref{sec:polBlogs}) 	     &  					&  					&  & 		& 		&  		 \\[0.2cm]			
			Human brain 			                 & \multirow{3}{*}{15.701} 	& \multirow{3}{*}{15.511} 	& \multirow{3}{*}{\textbf{15.412}} & \multirow{3}{*}{15.440}		& \multirow{3}{*}{15.663}		& \multirow{3}{*}{15.694}		 \\ 
			functional coactivations                 & & & & 		&		&  		 \\
			(Section~\ref{sec:brain})                 & & & & 		&		&  		 \\
			\toprule
			$K$                                        & 5		& 6		& 7		& 8		& 9		& 10 		\\
			\midrule \midrule
			Military alliances 		                 & \multirow{2}{*}{3.026$^*$}					    & \multirow{2}{*}{2.951$^*$} 					& \multirow{2}{*}{\textbf{2.885}$^*$} & \multirow{2}{*}{2.987$^*$}		& \multirow{2}{*}{3.113$^*$}		& \multirow{2}{*}{3.109$^*$} 		 \\
			(Section~\ref{sec:alliances})                 & & & & 		&		&  		 \\
			\bottomrule
		\end{tabular*}	
	\end{center}
	\caption{Resulting values of criterion (\ref{eq:modelSelect}) for all application networks considered in Section~\ref{sec:evalu}. 
		Specification refers to the factor of $10^4$ 
		($^*$or $10^3$). The lowest value per network is highlighted in bold.}
	\label{tab:crit}
\end{table}
This suggests choosing the correct number of three communities. 
The corresponding results of the estimation procedure with $K=3$ are illustrated in Figure~\ref{fig:synGraphon}. 
It can be clearly seen that the resulting \mbox{SBSGM} estimate (top right panel) promisingly captures the structure of the true model (top left). 
In line with this, comparing the estimated node positions 
with the true simulated ones (bottom right) shows that the latent quantities are appropriately recovered.  
More precisely, it exhibits that all three truly underlying groups are clearly separated and, in addition, also the within-community positions are well replicated. 
Altogether, the underlying structure can be precisely uncovered.

\subsubsection{Core-Periphery Structure}
\label{sec:hubNet}
\begin{figure}%
	\centering	
	\begin{minipage}[t]{0.625\textwidth}
		\centering
		\begin{minipage}[t]{0.48\textwidth}
			\centering
			\includegraphics[width=1\textwidth,trim={1.4cm 0.7cm 1.0cm 0.3cm},clip]{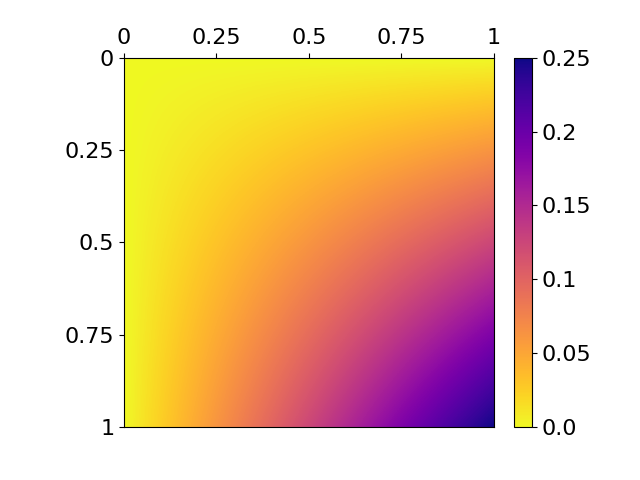}
		\end{minipage}
		\hfill
		\begin{minipage}[t]{0.48\textwidth}
			\centering
			\includegraphics[width=1\textwidth,trim={1.4cm 0.7cm 1.0cm 0.3cm},clip]{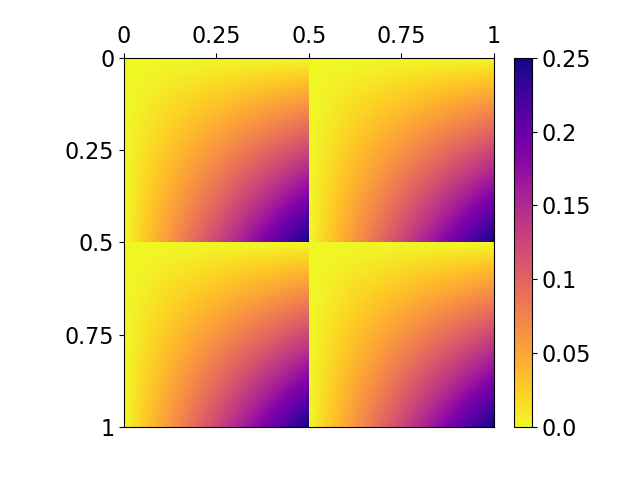}
		\end{minipage}
		\begin{minipage}[t]{0.48\textwidth}
			\centering
			\includegraphics[width=1\textwidth,trim={1.4cm 1cm 1.0cm 0.0cm},clip]{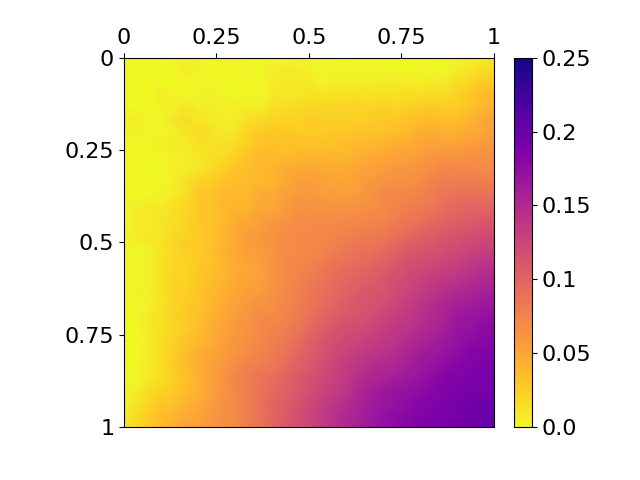}
		\end{minipage}
		\hfill
		\begin{minipage}[t]{0.48\textwidth}
			\centering
			\includegraphics[width=1\textwidth,trim={1.4cm 1cm 1.0cm 0.0cm},clip]{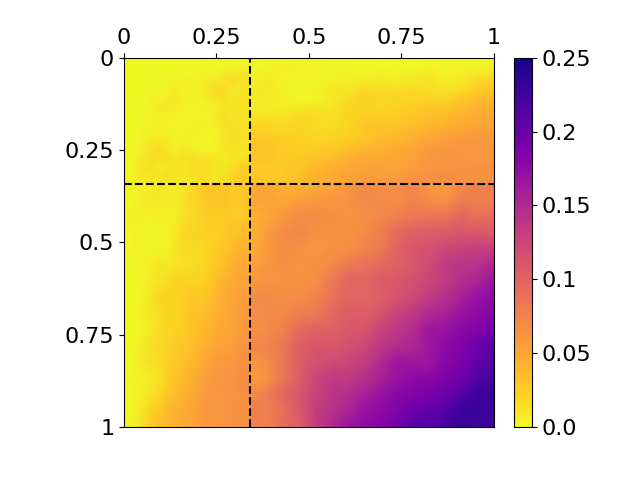}
		\end{minipage}
	\end{minipage}
	\begin{minipage}[c]{0.03\textwidth}
		\centering
		\vspace*{0.55cm}
		
		\includegraphics[width=.6\textwidth]{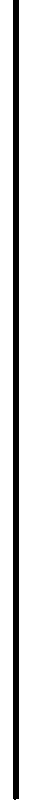}
	\end{minipage}
	\begin{minipage}[t]{0.3\textwidth}
		\centering
		\begin{minipage}[t]{1\textwidth}
			\centering
			\includegraphics[width=1\textwidth,trim={1.4cm 0.7cm 1.0cm 0.3cm},clip]{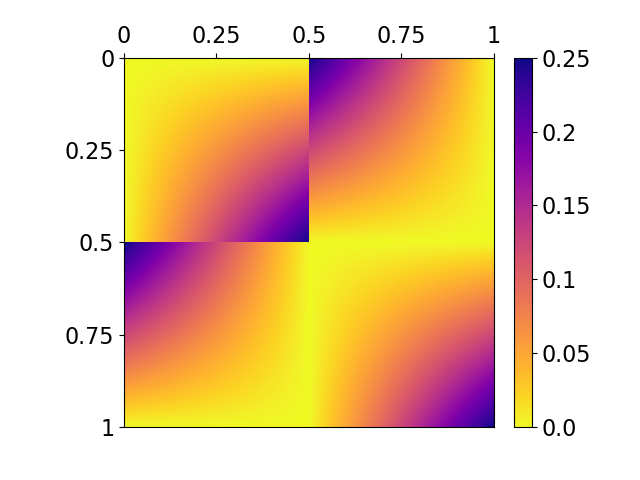}
		\end{minipage}
		\begin{minipage}[t]{1\textwidth}
			\centering
			\includegraphics[width=1\textwidth,trim={1.4cm 1cm 1.0cm 0.0cm},clip]{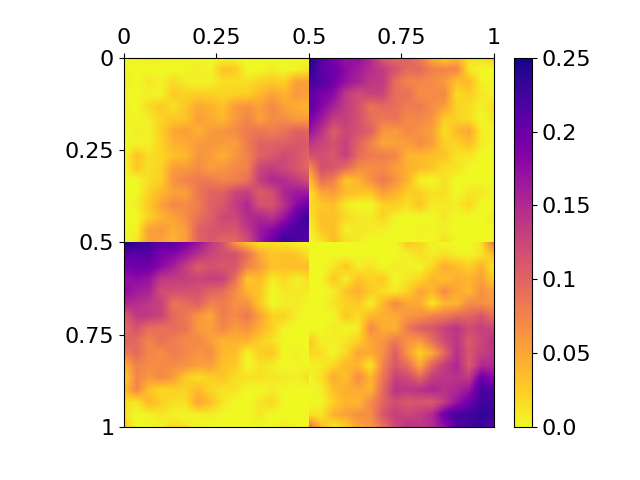}
		\end{minipage}
	\end{minipage}
	\caption{Estimation results for two synthetic \mbox{SBSGMs} (upper row). For the left model, two different representations with different numbers of groups are illustrated. 
	The estimates with the number of groups adopted from the above model representations are illustrated in the lower row. 
	The estimation is based on simulated networks of size $N = 500$.}
	\label{fig:synGraphon2}
\end{figure}

As a second simulation example, we consider the model in the top left plot of Figure~\ref{fig:synGraphon2}, which, at that,  
is equivalent to the one from the top middle plot. 
Apparently the ``true'' number of groups here is $K=1$, meaning that the left model is preferred. 
Nevertheless, both models describe the same structure and the estimation procedure should obviously follow only one representation, regardless of the chosen number of groups. 
To demonstrate the proceeding of our algorithm, we fit the model with both settings, $K=1$ and $K=2$. The results for simulated networks of size $N=500$ are illustrated in the lower row of the left-hand side of Figure~\ref{fig:synGraphon2}. 
This shows that both estimates follow the ``single-community'' representation, demonstrating the method's intuition of merging similar nodes. 
Additionally, comparing the estimates in terms of minimizing criterion (\ref{eq:modelSelect}) (see second row of Table~\ref{tab:crit}), the model fit with $K=1$ appears preferable over the one with $K=2$.

\subsubsection{Mixture of Assortative and Disassortative Structures under Equal Overall Attractiveness}
\label{sec:diffPrefNet}
We now amend the previous situation with regard to the ``two-community'' representation in the following spirit. 
Nodes which are highly connected within their own group now 
should only be poorly connected into the respective other community and vice versa. This leads us to the \mbox{SBSGM} represented in the top right plot of Figure~\ref{fig:synGraphon2}. More precisely, in this model, all nodes have the same expected degree, where nodes being weakly connected within their own community compensated their lack of attractiveness by reaching out to members of the respective other community. The structure of this \mbox{SBSGM} can clearly not be collapsed to a ``single-community'' representation. More importantly, considering such a structure from the SBM perspective, which, for the same $K$, inherently assumes a lower complexity, it also cannot be captured by degree correction. 
However, 
the \mbox{SBSGM} estimate at the bottom right shows that 
also in such a case, our algorithm is able to fully capture the underlying structure. 
Note that applying criterion (\ref{eq:modelSelect}) (see third row of Table~\ref{tab:crit}) actually yields the group number of $K=1$. This might be caused by the fact that, indeed, the structural break at $0.5$ goes only halfway through. In addition, the decision is quite close compared to the setting of $K=2$.

\subsection{Real-World Networks}
\label{sec:real-world}
For evaluating our method with regard to real-world examples, we consider three networks from different domains, comprising social/political sciences and neurosciences. Besides their different domains, the networks differ in their inherent structure, including the overall density. An overview of the networks' most relevant coefficients is given in Table~\ref{tab:nets}. 
\begin{table}%
	\begin{center}
		\begin{tabular*}{\textwidth}{l @{\extracolsep{\fill}} ccc}
			\toprule
									 & Number of nodes 		    & Average degree 			& Overall density \\
			\midrule
			Political blogs 	     & 1222 					& 27.31 					& 0.022 \\[0.2cm]
			Military alliances 		 & 141 					    & 24.16 					& 0.173 \\ [0.2cm]
			Human brain 			 & \multirow{2}{*}{638} 	& \multirow{2}{*}{58.39} 	& \multirow{2}{*}{0.092} \\ 
			functional coactivations & & & \\
			\bottomrule
		\end{tabular*}
	\end{center}
	\caption{Details about real-world networks used as application examples.}
	\label{tab:nets}
\end{table}

\subsubsection{Political Blogs}
\label{sec:polBlogs}
\begin{figure}%
		\centering
		\begin{minipage}[t]{1\textwidth}
			\centering
			\includegraphics[width=.55\textwidth,trim={1cm -.1cm 1cm 1.2cm},clip]{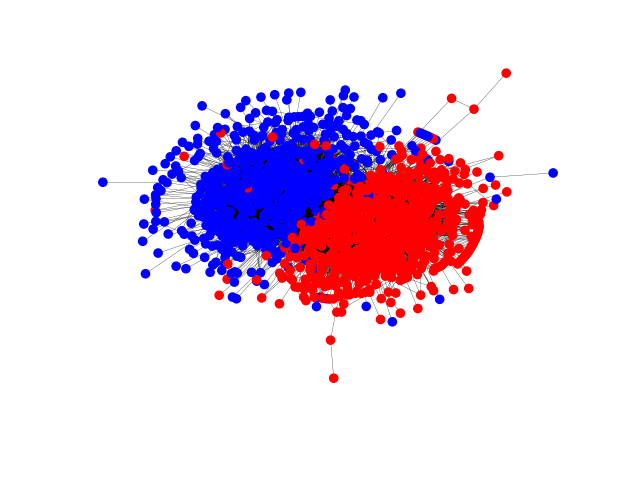}
		\end{minipage}
		\vspace*{-1.5cm}
		
		\begin{minipage}[t]{0.49\textwidth}
			\centering
			\includegraphics[width=1\textwidth]{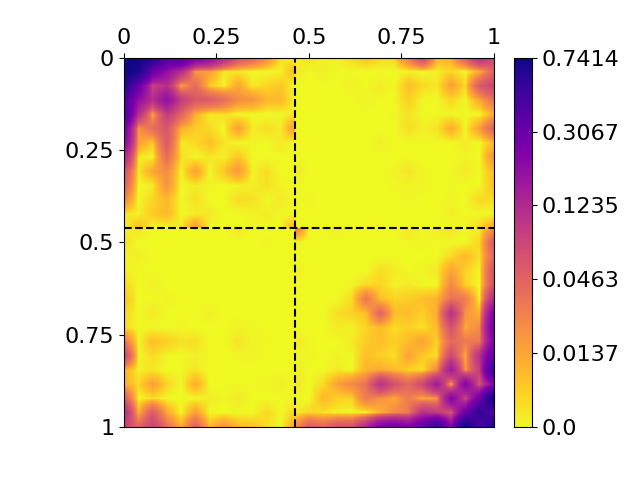}
		\end{minipage}
		\hfill
		\begin{minipage}[t]{0.49\textwidth}
			\centering
			\includegraphics[width=1\textwidth,trim={1cm -.1cm 1cm 1.2cm},clip]{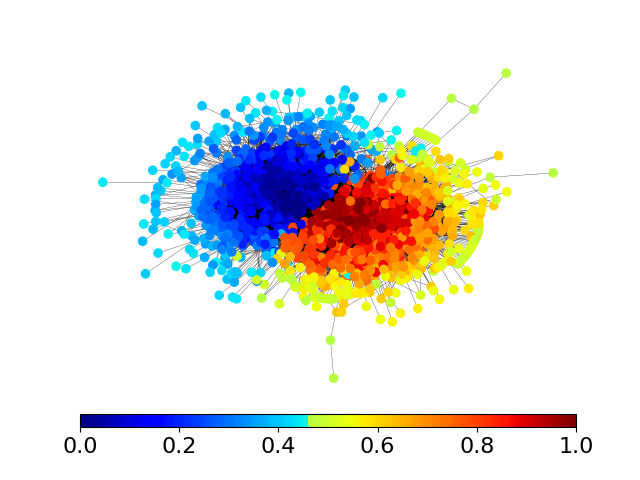}
		\end{minipage}
		\vspace*{-0.4cm}
		
		\begin{minipage}[t]{0.49\textwidth}
			\centering
			\hspace*{-.61cm}
			\includegraphics[width=1.001\textwidth]{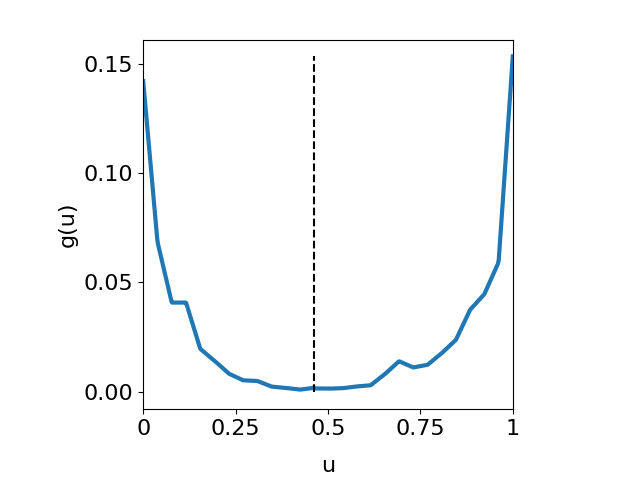}  %
		\end{minipage}
		\hfill
		\begin{minipage}[t]{0.49\textwidth}
			\centering
			\includegraphics[width=1\textwidth,trim={1cm -.1cm 1cm 1.2cm},clip]{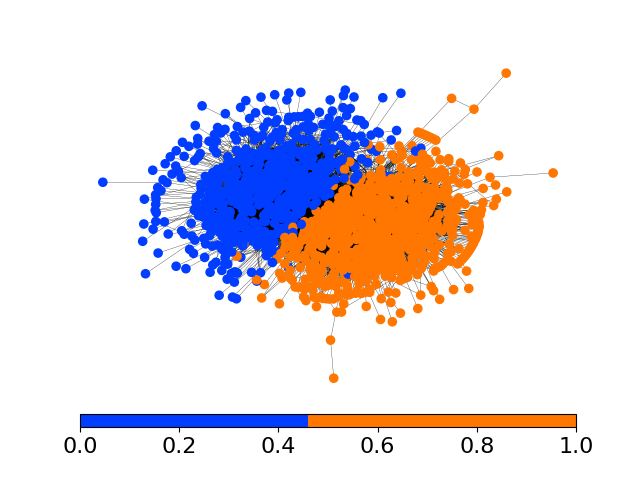}
		\end{minipage}
	\vspace*{0.15cm}
	\caption{\mbox{SBSGM} estimation for the political blog network (top plot, with blue for `liberal' and red for `conservative'). The estimated \mbox{SBSGM} (in log scale) is depicted 
	at the middle left. 
	The plot at the bottom left illustrates the corresponding marginal function 
	$\hat{g}_{\bzeta}(u) = \int \hat{w}_{\bzeta}(u,v) \diff v$ 
	for $u \in [0,1]$. 
	In the right column of the two lower rows, 
	the network is shown with coloring referring once to the node positions $\hat{U}_i$ (top) and once to the derived community memberships (bottom).}
	\label{fig:polBlogs}
\end{figure}

The political blog network has been assembled by \cite{Adamic2005} and consists of 1222 nodes (after extracting the largest connected component). 
The network's nodes represent political blogs of which $586$ are liberal and $636$ are conservative, according to manual labeling \citep{Adamic2005}. 
Here, an edge between two blogs illustrates a web link pointing from one blog to the other within a single-day snapshot in 2005. 
For our purpose, these links are interpreted in an undirected fashion. 
The arising network with political labels included 
is illustrated in the top plot of Figure~\ref{fig:polBlogs}. 
Note that the exact same network has also been used by \cite{Karrer2011} for demonstrating the enhancement achieved through their degree-corrected variant of the SBM. 
For our method, we again start with determining the number of groups. In accordance with the number of political orientations, criterion (\ref{eq:modelSelect}) suggests to set $K=2$ (see fourth row of Table~\ref{tab:crit}). 
The corresponding results are illustrated at the two lower rows of Figure~\ref{fig:polBlogs}. 
The predicted group assignments depicted at the bottom right exhibit a clear separation and, moreover, show a broad concordance with the manually assigned labels. 
This is also reflected in a similar size ratio of $565$ (mostly liberals) to $657$ (mostly conservatives). In addition to the pure community memberships, with our method we also 
gain information about 
the within-community positions. These are visualized by the middle right plot, 
revealing additional local structures within the network. That is, for example, a community-wise division into core and periphery nodes, where such a core-periphery structure is a well-known phenomenon in the linkage within the World Wide Web. The \mbox{SBSGM} estimate 
and the corresponding marginal function according to (\ref{eq:gMono}) 
(depicted in the middle and bottom left plot, respectively) 
further indicate the presence of hubs, meaning a minority subgroup of nodes that are much more densely connected than others. This can be 
deduced from 
the narrow intense regions in the \mbox{SBSGM} and the 
steep slopes 
in the 
marginal function. 
Moreover, 
the \mbox{SBSGM} reveals a domination of assortative structures because the overall intensity within the two communities is much higher than between them. 
Altogether, we gain profound information about the structure within the network.

\subsubsection{Military Alliances}
\label{sec:alliances}
\begin{figure}%
	\centering
	\begin{minipage}[t]{0.49\textwidth}
		\centering
		\includegraphics[width=1\textwidth]{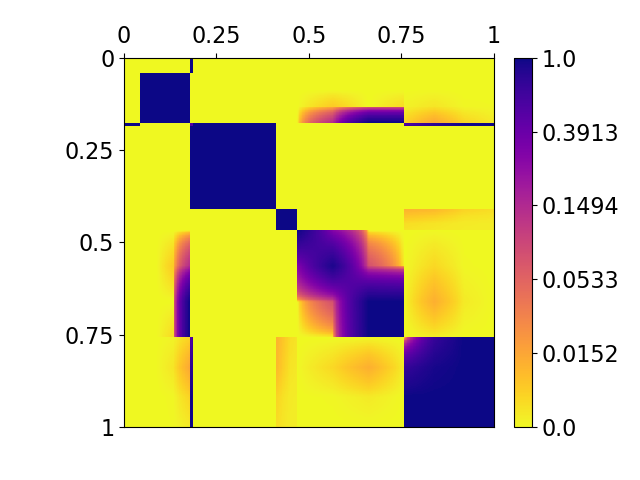}  %
	\end{minipage}
	\begin{minipage}[t]{0.49\textwidth}
		\centering
		\includegraphics[width=1\textwidth]{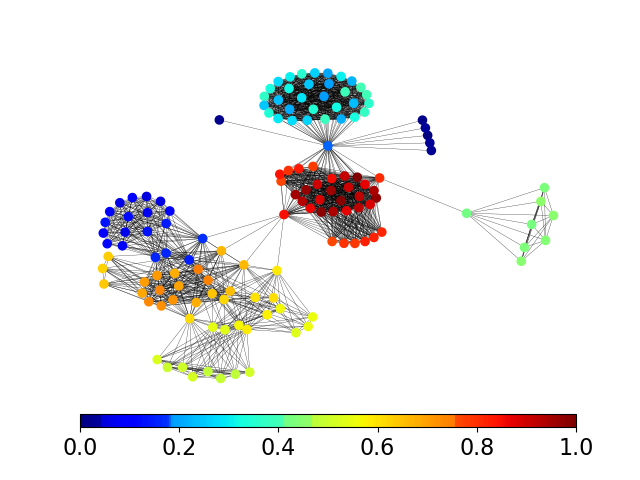}  %
	\end{minipage}
	\vspace*{-0.1cm}
	
	\begin{minipage}[t]{1\textwidth}
		\centering
		\includegraphics[width=1\textwidth, trim={0 3.2cm 0 2.5cm}, clip]{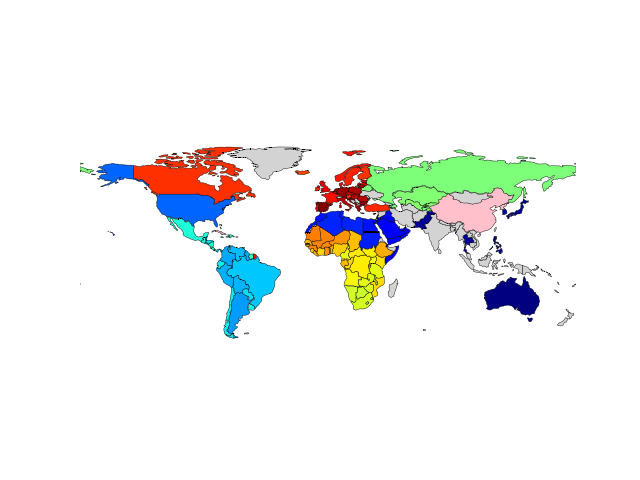}  %
	\end{minipage}
	\vspace*{-.9cm}
	
	\begin{minipage}[t]{1\textwidth}
		\centering
		\includegraphics[width=1\textwidth, trim={0 3.2cm 0 2.5cm}, clip]{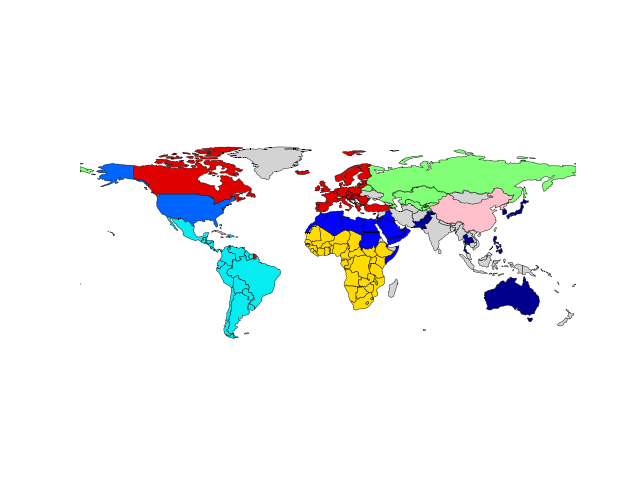}  %
	\end{minipage}
	\caption{\mbox{SBSGM} estimation for the military alliance network. The estimated \mbox{SBSGM} (in log scale) and the network with node coloring referring to 
	$\hat{U}_i \in [0,1]$ 
	are depicted at the top left and top right, respectively. The two lower plots show the world map with colors indicating the exact position in the \mbox{SBSGM} (middle) and the resulting community membership (bottom). China, Cuba, and North Korea (colored in pink) form an isolated group and therefore have been excluded from the estimation procedure. Countries which do not appear in the data set and hence are assumed not to have any strong military alliance are colored in gray.}
	\label{fig:milAll}
\end{figure}

As a second real-world example, we consider the military alliances among the
world’s nations. These data have been gathered and are provided by the Alliance Treaty Obligations and Provisions project (\citealp{Leeds2002}). More specifically, from the available data, we extracted only the strong military alliances which were lately in force. That means, an edge between two countries is included if they have a current agreement in the form of an offensive or a defensive pact. Such a pact would force the one country to militarily intervene when the other one has come into an offensive or defensive military conflict. This network, which, referring to criterion (\ref{eq:modelSelect}), decomposes into seven communities (see last row of Table~\ref{tab:crit}), is shown in the top right plot of Figure~\ref{fig:milAll}. 
With respect to the network formation, the estimated node positions (depicted by node coloring) appear reasonable, which involves both the group assignment and the within-community location. 
The \mbox{SBSGM} estimate, which is shown in the top left plot, reveals again a very dominant assortative structure. However, there are few groups which also have a strong connection to other groups. Transferring 
the node positions and the resulting community memberships 
to the world map, as shown in the 
two lower plots,  
allows to deduce certain political structures and relations. Regarding the communities (bottom plot), it can be seen that almost all of them consist exclusively of neighboring countries, implying that those arrange similar strong military alliances. In combination with the discovered assortative structure, one can additionally conclude that countries which are geographically close are likely to form a military alliance. Furthermore, the within-community positions can be consulted to gain additional insight into the local structure (see middle plot). For example, considering the countries of the central and southern part of Africa (yellow community in the bottom plot), it can be seen that there is a more or less stringent transition from Southern Africa via Central/East Africa to West Africa.

\subsubsection{Human Brain Functional Coactivations}
\label{sec:brain}
\begin{figure}%
	\centering
	\begin{minipage}[t]{0.49\textwidth}
		\centering
		\includegraphics[width=1\textwidth]{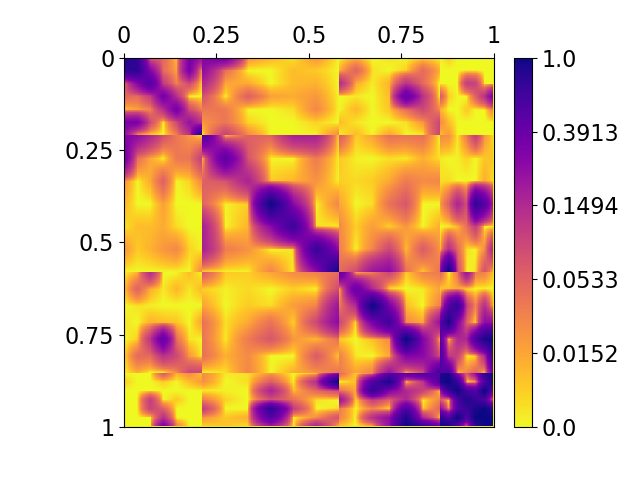}
	\end{minipage}
	\hfill
	\begin{minipage}[t]{0.49\textwidth}
		\centering
		\includegraphics[width=1\textwidth]{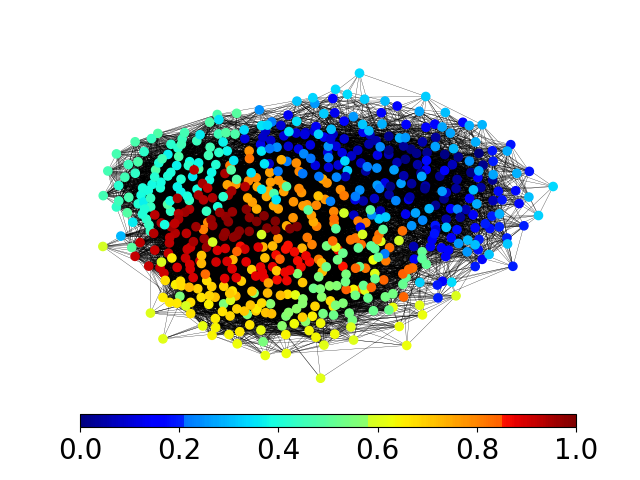}
	\end{minipage}
	
	\begin{minipage}[t]{0.32\textwidth}
		\centering
		\includegraphics[width=.98\textwidth,trim={1.5cm 0.5cm 1.5cm 0.5cm},clip]{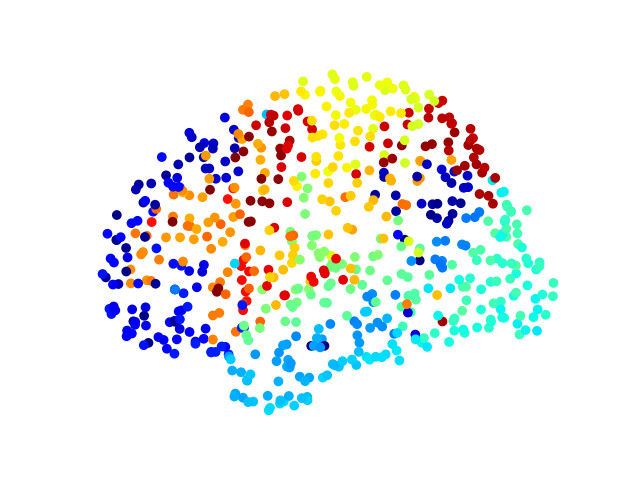}
	\end{minipage}
	\begin{minipage}[t]{0.32\textwidth}
		\centering
		\includegraphics[width=.98\textwidth,trim={1.5cm 0.5cm 1.5cm 0.5cm},clip]{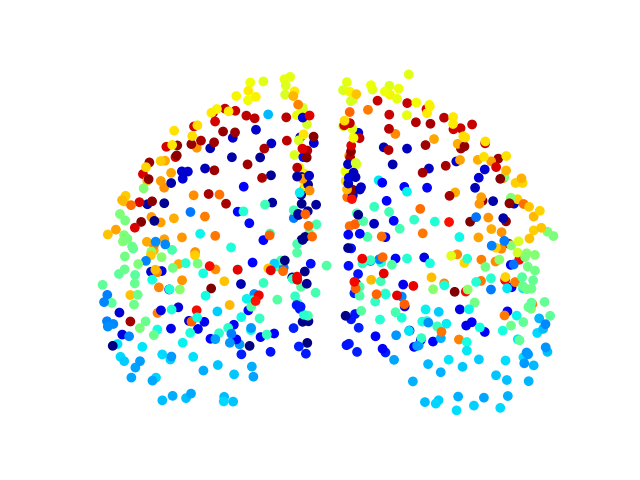}
	\end{minipage}
	\begin{minipage}[t]{0.32\textwidth}
		\centering
		\includegraphics[width=.98\textwidth,trim={1.5cm 0.5cm 1.5cm 0.5cm},clip]{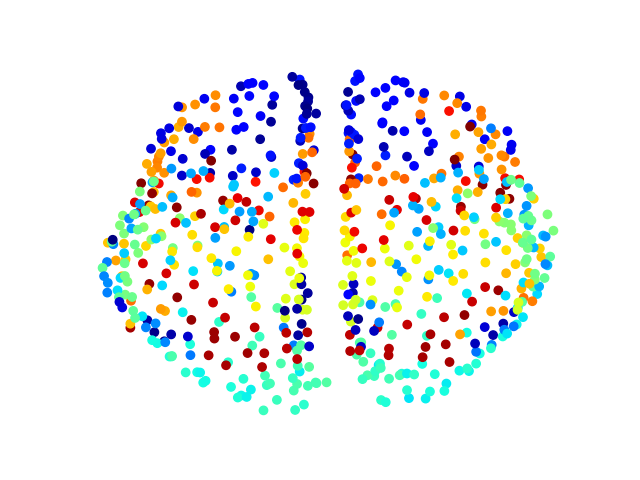}
	\end{minipage}	
	\begin{minipage}[t]{0.32\textwidth}
		\centering
		\includegraphics[width=.98\textwidth,trim={1.5cm 0.5cm 1.5cm 0.5cm},clip]{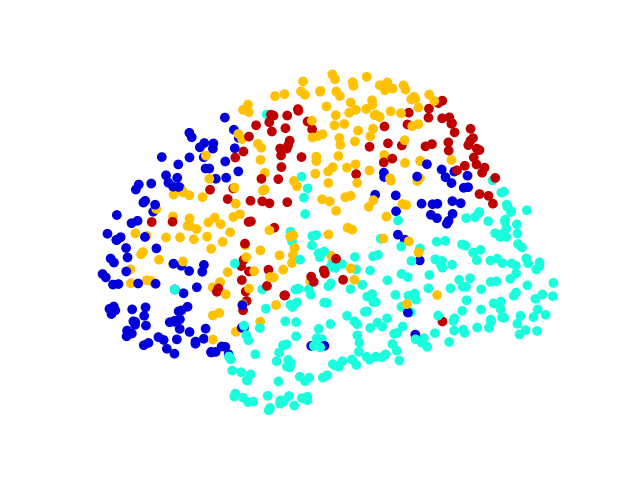}
	\end{minipage}
	\begin{minipage}[t]{0.32\textwidth}
		\centering
		\includegraphics[width=.98\textwidth,trim={1.5cm 0.5cm 1.5cm 0.5cm},clip]{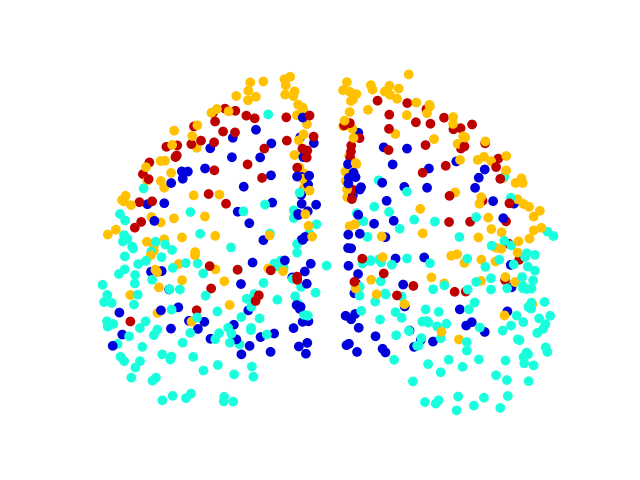}
	\end{minipage}
	\begin{minipage}[t]{0.32\textwidth}
		\centering
		\includegraphics[width=.98\textwidth,trim={1.5cm 0.5cm 1.5cm 0.5cm},clip]{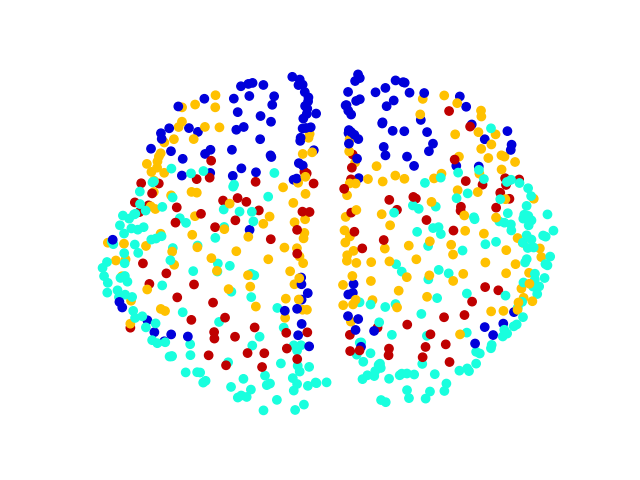}
	\end{minipage}
	\caption{\mbox{SBSGM} estimation for the human brain functional coactivation network (top right, with coloring referring to 
	$\hat{U}_i \in [0,1]$). 
	The \mbox{SBSGM} estimate %
	(in log scale) is depicted at the top left. The lower six plots show the local positions of the human brain regions in anatomical space with coloring referring to $\hat{U}_i \in [0,1]$ (middle row) and derived community memberships (bottom row). The different perspectives are side view (left column), front view (middle column), and top view (right column).
	}
	\label{fig:brain}
\end{figure}

We conclude the real-world data examples by considering the human brain functional coactivation network. This network is accessible thorough the Brain Connectivity Toolbox (\citealp{brain:10}) and has been assembled by \cite{Crossley2013} via meta-analysis. 
More precisely, the provided weighted network matrix represents the ``estimated [\ldots] similarity (Jaccard index) of the activation patterns across
experimental tasks between each pair of 638 brain regions'' (\citealp{Crossley2013}), where this similarity is additionally ``probabilistically thresholded''. From that, we construct an unweighted graph by including a link between all pairs of brain regions which have a significant similarity, meaning a positive score in the original data. 
For the arising network, determining the number of groups using criterion (\ref{eq:modelSelect}) yields three communities (see last-but-one row of Table~\ref{tab:crit}). However, since the decision seems tight and \cite{Crossley2013} choose a regular SBM with four communities for fitting the data, we also here choose $K=4$ 
to allow for comparison. 
The corresponding estimation results of the algorithm applied to this network are illustrated in Figure~\ref{fig:brain}. 
Also here, the \mbox{SBSGM} estimate in the top left plot reveals an assortative structure, though less pronounced. Apparently, there are several pairs of node bundles which, according to the latent space, are not close together but still well connected. This generally means that connectedness not necessarily needs to be accompanied by similar behavior. 

Considering the network at top right, it reveals that the node positioning and clustering is in line with the network's formation. 
Transferring these results to the anatomical space, as is done in the two lower rows, provides information about the relation between similar behavior and anatomical location. At that, the areas of all four found communities (bottom row) can be clearly delimited, although these areas are not always solidly connected. Besides, they seem to have a rather specific shape. For example, the blue community spreads out over the front part of the frontal lobe and to some extent over the rear part of the parietal lobe. In contrast, the cyan group occupies more the temporal lobe and the occipital lobe. In addition, on the basis of the within-community positions (middle row), one here can see that the latter community subdivides into those two lobes. 

Altogether, it can be demonstrated that our novel modeling approach is very flexible when it comes to capturing the structure within complex networks. 
This, in combination with its favorable interpretability---provided through group assignments and within-community positions---, makes it 
a helpful tool to gain further insight and 
to draw more profound 
scientific conclusions.

\section{Discussion and Conclusion}
\label{sec:discuss}
Despite their close relationship, the stochastic blockmodel and the (smooth) graphon model 
have mainly been developed separately until now. 
To address this shortcoming, the paper aimed at combining both model formulations to develop a novel modeling approach that unifies and, more importantly, generalizes the two detached concepts. 
The resulting stochastic block smooth graphon model consequently unites the individual capabilities of the two underlying approaches. 
That are, on the one hand, clustering the network and, on the other hand, including local structure in the form of smooth differences within communities. 
Moreover, utilizing previous results on SBM and SGM estimation, we presented an EM-type algorithm for reliably estimating the semiparametric model formulation. 

Although the \mbox{SBSGM} arises from combining 
the SBM and the SGM, 
connections 
to other statistical network models can be established. This is further elaborated in the Supplementary Material. For example, an approximation of the latent distance model (\citealp{Hoff2002}) can be formulated for any dimension by appropriately partitioning and mapping the latent space to the unit interval as the support of the \mbox{SBSGM}. 
Moreover, the \mbox{SBSGM} is also able to cover to some extent the structure of the degree-corrected SBM (\citealp{Karrer2011}) in a natural way. This can be accomplished by 
restricting the slices $w_{\bzeta}(u,\cdot)$ to be proportional within communities, i.e.\ setting $w_{\bzeta}(u,\cdot) \equiv c \cdot w_{\bzeta}(u^\prime,\cdot)$ with $c \in \mathbb{R}_+$ for $u, u^\prime \in [\zeta_{k-1}, \zeta_k)$. Such a specification implies an equivalent connectivity behavior with different attractiveness. 
The applicability to such a situation has been demonstrated by the political blog example of Section~\ref{sec:polBlogs}. Finally, we argue that---from a conceptional perspective---the \mbox{SBSGM} is also related to the hierarchical exponential random graph model developed by \cite{Schweinberger2015}. This is because in both models, the set of nodes is divided into ``neighborhoods'' (\citealp{Schweinberger2015}) within which the local structure is then modeled by a further approach. In the HERGM, this local structure is captured in the form of exponential random graph models, while in the \mbox{SBSGM}, this is known to be done using SGMs. 
In this regard, a connection between the graphon model and the ERGM has been elaborated by \cite{Chatterjee2013}, \cite{Yin2016}, and \cite{Krioukov2016}. 

Besides its theoretical capabilities and connections, we demonstrated the practical applicability of the \mbox{SBSGM} with reference to both simulated and real-world networks. 
The estimation results of Section~\ref{sec:evalu} clearly revealed that our novel modeling approach of clustering nodes and parallel including smooth structural differences is able to capture various types of complex structural patterns. Overall, the \mbox{SBSGM} is a very flexible tool for modeling complex networks, which, as such, helps to uncover the network's structure in more detail and thus enables to get a better understanding of the 
underlying processes.

\section*{Acknowledgment}

The project was partially supported by the European Cooperation in Science and Tech\-nology [COST Action CA15109 (COSTNET)]. 

\noindent
This research did not receive any specific grant from funding agencies in the public, commercial, or not-for-profit sectors.

\noindent
Declarations of interest: none.

\bigskip
\begin{center}
	{\large\bf SUPPLEMENTARY MATERIAL}
\end{center}
The Supplementary Material comprises elaborations about intuition and justification of the EM-type algorithm presented in the paper, as well as the concrete formulation of links to other models. 
Moreover, we have implemented the EM-based estimation routine described in the paper
in a free and open source \texttt{Python} package, which is publicly available on https://github.com/BenjaminSischka/SBSGMest.git (\citealp{Sischka:22b}).

\clearpage

\bibliographystyle{chicago}

\bibliography{/home/bens/Documents/Mendeley/MixtureGraphonModel2.bib}

\appendix
\section*{Appendix}
\addcontentsline{toc}{section}{Appendices}
\renewcommand{\thesubsection}{\Alph{subsection}}

\subsection*{The Gibbs Sampling of Node Positions and Subsequent Adjustments}
\label{AppSec:gibbs}
In the EM-type algorithm presented in the paper, we apply the Gibbs sampler in the E-step to achieve appropriate node positions conditional on $\bY=\by$ and given $w_{\bzeta}(\cdot,\cdot)$. 
That means, we aim to stepwise update the $i$-th component of the current state of the Markov chain, $\bU^{<t>} = (U^{<t>}_{1},\ldots,U^{<t>}_{N})$. This 
is done by setting $U^{<t+1>}_{j} := U^{<t>}_{j}$ for $j \neq i$ and for $U^{<t+1>}_i$ drawing from the full-conditional distribution as formulated in (\ref{eq:Gibb1}) of the paper. 
To do so, we make use of a mixture proposal which differentiates between 
remaining within and switching the community. 
This is appropriate due to different structural relations with respect to $U^{<t>}_i$, where nearby positions within the same community imply similar connectivity patterns. 
To this end, 
we split the proposing procedure into two separate steps. First, we randomly choose the proposal type, i.e.\ either remaining within or switching the community. This is done by drawing from a Bernoulli distribution with $\nu \in [0,1]$ as the probability of remaining within group. 
Secondly, conditional on the proposal type, we either draw from $[\zeta_{k_i-1}, \zeta_{k_i})$ or from $[0,1] \setminus [\zeta_{k_i-1}, \zeta_{k_i})$, where $k_i \in \{1,\ldots,K\}$ is the community including $U^{<t>}_i$, i.e.\ $U^{<t>}_i \in [\zeta_{k_i-1}, \zeta_{k_i})$. 
For a proposal within the current community, we employ a normal distribution under a compressed logit link. 
To be precise, we first define $V_i^{<t>} = \log \{ (U_i^{<t>} - \zeta_{k_i-1}) / (\zeta_{k_i} - U_i^{<t>}) \}$, then we draw $V_i^*$ from $\text{Normal}(V_i^{<t>}, \sigma^2)$ with an appropriate value for the variance $\sigma^2$,  %
and finally we calculate $U_i^* = \{\exp(V_i^*) / (1 + \exp(V_i^*))\} \cdot (\zeta_{k_i} - \zeta_{k_i-1}) + \zeta_{k_i-1}$. Consequently, for $U_i^* \in [\zeta_{k_i -1}, \zeta_{k_i})$, 
the proposal density follows
\begin{multline*}
p(U_i^* \mid U_i^{<t>} ) \propto \nu \cdot \frac{1}{(U_i^{*} - \zeta_{k_i-1}) (\zeta_{k_i} - U_i^{*})} \cdot \exp \left\{ - \frac{1}{2 \sigma^2} \left[ \vphantom{\log \left\{ (U_i^{<t>} - \zeta_{k_i-1}) / (\zeta_{k_i} - U_i^{<t>}) \right\}} \right. \right. \log \left\{ (U_i^* - \zeta_{k_i-1}) / (\zeta_{k_i} - U_i^*) \right\} \\ 
- \log \left\{ (U_i^{<t>} - \zeta_{k_i-1}) / (\zeta_{k_i} - U_i^{<t>}) \right\} \left. \vphantom{\frac{1}{2 \sigma^2}} \left. \vphantom{\log \left\{ (U_i^{<t>} - \zeta_{k_i-1}) / (\zeta_{k_i} - U_i^{<t>}) \right\}} \right]^2 \right\},
\end{multline*}
yielding a ratio of proposals in the form of 
\[
\frac{p(U_i^{<t>} \mid U_i^* )}{p(U_i^* \mid U_i^{<t>} )} = \frac{(U_i^{*} - \zeta_{k_i-1}) (\zeta_{k_i} - U_i^{*})}{(U_i^{<t>} - \zeta_{k_i-1}) (\zeta_{k_i} - U_i^{<t>})}.
\]
Regarding the proposal under switching the community, no information about the relation to $U^{<t>}_i$ is given beforehand. Hence, in this case, 
we apply a uniform proposal restricted to the segments of all other groups. To be precise, for $U_i^*$ we draw from a uniform distribution with the support $[0, \zeta_{k_i-1}) \cup [\zeta_{k_i}, 1]$. This means that the proposal density is given as $p(U_i^* \mid U_i^{<t>} ) = (1- \nu) / (1 - (\zeta_{k_i} - \zeta_{k_i-1})) \cdot \indFun{U_i^* \in [0, \zeta_{k_i-1}) \cup [\zeta_{k_i}, 1]}$, 
yielding 
for $U^{*}_i \in [\zeta_{k_i^*-1}, \zeta_{k_i^*})$ with $k_i^* \neq k_i$
a proposal ratio of 
\[
\frac{p(U_i^{<t>} \mid U_i^* )}{p(U_i^* \mid U_i^{<t>} )} = \frac{1 - (\zeta_{k_i} - \zeta_{k_i-1})}{1 - (\zeta_{k_i^*} - \zeta_{k_i^*-1})}.
\] 

Having defined the proceeding for proposing a new position for node $i$, including the calculations of the corresponding density ratios, 
we are now able to specify the acceptance probability. Hence, we accept the proposed value and therefore set $U_i^{<t+1>} := U^{*}_i$ with a probability of
\begin{align*}
\min \left\{ 1, \quad \prod_{j \neq i} 
\vphantom{\left(  \frac{(1-w_{\bzeta} (U^*_j, U^{<t>}_{j}))}{(1-w_{\bzeta}(U^{<t>}_{j}, U^{<t>}_{j}))}   \right)^{1-y_{ji}}} \right.  \left[ \left(  \frac{w_{\bzeta}(U_i^*, U^{<t>}_{j})}{w_{\bzeta}(U^{<t>}_{i}, U^{<t>}_{j})}  \right)^{y_{ij}}   
\left(  \frac{1-w_{\bzeta}(U^*_i, U^{<t>}_{j})}{1-w_{\bzeta}(U^{<t>}_{i}, U^{<t>}_{j})}   \right)^{1-y_{ij}} \right] \frac{p(U_i^{<t>} \mid U_i^* )}{p(U_i^* \mid U_i^{<t>} )} \left. \vphantom{\left(  \frac{(1-w_{\bzeta}(U^*_j, U^{<t>}_{j}))}{(1-w_{\bzeta}(U^{<t>}_{j}, U^{<t>}_{j}))}   \right)^{1-y_{ji}}}
\right\}.
\end{align*}
If we do not accept $U_i^*$, we set $U^{<t+1>}_{i} := U^{<t>}_{i}$. The consecutive 
drawing and updating of the components 
$U_1,\ldots,U_N$ then provides a proper Gibbs sampling sequence. 
After cutting the burn-in phase and appropriate thinning, calculating the sample mean of the simulated values consequently yields an approximation of the marginal conditional mean $\Ev(U_i \mid \by)$. 
To be precise, for appropriately estimating $U_i$ in the $m$-th iteration of the EM algorithm, we define
\begin{align}
\hat{U}_i^{(m)} = \frac{1}{n} \sum_{s=1+b}^{n+b} U^{<s \cdot N \cdot r>}_{i},
\label{eq:gibbsMean}
\end{align}
where $b \in \mathbb{N}$ represents a burn-in parameter, $r \in \mathbb{N}$ describes a thinning factor, and $n$ is the number of MCMC states which are taken into account. 

However, as discussed in Section~\ref{subsec:estep} of the paper, these estimates need to be further adjusted in a two-fold manner, which also includes adjusting the community boundaries. 
Starting with Adjustment~1, we relocate the boundaries $\zeta_k$ such that the group allocations correspond to the proportions of the realized groups, 
meaning we set
\[
\hat{\zeta}_k^{(m+1)} = \frac{\sum_i \indFun{\hat{U}_i^{(m)} < \hat{\zeta}_k^{(m)}}}{N}. 
\]
Note that this calculation represents an estimate of the transformation $\varphi_1^\prime (\hat{\zeta}_k^{(m)})$ 
with $\hat{\varphi}_1^\prime (\cdot)$ as described in Section~\ref{subsec:estep} of the paper. 
In fact, it is advisable to make small adjustments in early iterations since, 
in the beginning, the result of the E-step is 
rather rough. 
We therefore make use of step-size adjustments in the form of
\[
\hat{\zeta}_{k}^{(m+1)}=\delta^{(m+1)} \frac{\sum_{i} \indFun{
		\hat{U}_i^{(m)}
		< \hat{\zeta}_{k}^{(m)}}}{N} + \left(1-\delta^{(m+1)}\right) \frac{k}{K}. 
\]
In this specification, the weighting $\delta^{(m+1)} \in[0,1]$ with $\delta^{(m+1)} \geq \delta^{(m)}$ induces a step-size adaptation from a priori equidistant boundaries to boundaries implied by observed frequencies. Such step-size adaptation is recommendable to prevent the community size to shrink too substantially before the structure of the community has been evolved properly. In 
general, $\delta^{(m+1)}$ is chosen to be one in the last iteration. This concludes Adjustment~1 with respect to the community boundaries. 

We proceed with applying Adjustment~1 and Adjustment~2 to the posterior means derived from 
expression (\ref{eq:gibbsMean}). 
To do so, we order all $\hat{U}_i^{(m)}$ in the original blocks by ranks and rescale them to the new blocks defined through $[\hat{\zeta}_{k-1}^{(m+1)}, \hat{\zeta}_{k}^{(m+1)})$. 
That means, we first assign 
communities through $\mathcal{C}_k^{(m)} = (i \in \{1,\ldots,N\}: \, \hat{\zeta}_{k-1}^{(m)} \leq 
\hat{U}_i^{(m)} 
< \hat{\zeta}_{k}^{(m)})$ 
with sizes $N_k^{(m)} = \vert \mathcal{C}_k^{(m)} \vert $. To enforce equidistant adjusted positions within the new community boundaries, we then calculate for all $j \in \mathcal{C}_k^{(m)}$
\begin{align}
\hat{U}_j^{\prime\prime (m)}
= \frac{\operatorname{rank}_{k}(
	\hat{U}_j^{(m)}
	)}{N_{k}^{(m)} +1} (\hat{\zeta}_{{k}}^{(m+1)} - \hat{\zeta}_{{k}-1}^{(m+1)}) + \hat{\zeta}_{{k}-1}^{(m+1)}
\label{eq:adjUs}
\end{align}
with $\operatorname{rank}_{k}(
\hat{U}_j^{(m)}
)$ being the rank from smallest to largest of the element $
\hat{U}_j^{(m)}
$ within all positions in community $k$, i.e.\ within the tuple $(
\hat{U}_i^{(m)}
: \, i \in \mathcal{C}_k^{(m)})$. These calculations, which represent an estimate of $\varphi_2^\prime \circ \varphi_1^\prime (\hat{U}_j^{(m)})$ with $\varphi_1^\prime (\cdot)$ and $\varphi_2^\prime ()$ as described in Section~\ref{subsec:estep} of the paper, are applied to all communities $k=1,\ldots,K$. This concludes applying Adjustment~1 and Adjustment~2 to the latent quantities.

\end{document}